# Gyrokinetic analysis and simulation of pedestals, to identify the culprits for energy losses using "fingerprints"

M. Kotschenreuther, X. Liu, D.R. Hatch, S. Mahajan, L. Zheng
University of Texas at Austin
Austin, USA
Email: mtk@austin.utexas.edu

A. Diallo
Princeton Plasma Physics Laboratory
Princeton, USA

R. Groebner and the DIII-D TEAM
General Atomics
San Diego, USA

J. C. Hillesheim, C. F. Maggi, C. Giroud, F. Koechl, V. Parail, S. Saarelma, E. Solano, and JET Contributors*
Culham Centre for Fusion Energy, Culham Science Centre
Abingdon, UK
*See the author list of X. Litaudon et al 2017 Nucl. Fusion 57 102001

A. Chankin
Max-Planck-Institut für Plasmaphysik
Garching bei München, Germany

**Abstract**

Fusion performance in tokamaks hinges critically on the efficacy of the Edge Transport Barrier (ETB) at suppressing energy losses. The new concept of "fingerprints" is introduced to identify the instabilities that cause the transport losses in the ETB of many of today's experiments, from widely posited candidates. Analysis of the Gyrokinetic-Maxwell equations, and gyrokinetic simulations of experiments, find that each mode type produces characteristic ratios of transport in the various channels: density, heat and impurities. This, together with experimental observations of transport in some channel, or, of the relative size of the driving sources of channels, can identify or determine the dominant modes causing energy transport. In multiple ELMy H-mode cases that are examined, these fingerprints indicate that MHD-like modes are apparently not the dominant agent of energy transport; rather, this role is played by Micro-Tearing Modes (MTM) and Electron Temperature Gradient (ETG) modes, and in addition, possibly Ion Temperature Gradient (ITG)/Trapped Electron Modes (ITG/TEM) on JET. MHD-like modes may dominate the electron particle losses. Fluctuation frequency can also be an important means of identification, and is often closely related to the transport fingerprint. The analytical arguments unify and explain previously disparate experimental observations on multiple devices, including DIII-D, JET and ASDEX-U, and detailed simulations of two DIII-D ETBs also demonstrate and corroborate this.

## I. Introduction

Ever since its experimental discovery, the H-mode [1], with its highly boosted energy confinement time ($\tau_E$), has been the centerpiece of nuclear fusion strategies in magnetically confined plasmas. High $\tau_E$, the primary requirement for fusion gain, is associated with the formation of an Edge Transport Barrier (ETB) that greatly impedes energy losses; the ETB (pedestal) is the defining feature of an H-mode configuration.

Understanding the physical processes that determine the transport of energy, particles and impurities in the pedestal is, naturally, crucial to 1) detailed interpretation of the current experiments, and 2) to reliably project and optimize H-modes for future burning plasmas. Clearly, one must begin with identifying the dominant transport agents in existing experimental pedestals.

To appreciate the larger context of the problem we set ourselves to investigate, let us consider for a moment, the importance of pedestal energy losses in determining the core energy confinement. Core profiles in H-mode are usually stiff. Stiffness implies that the profiles are weakly dependent on heating power but strongly dependent on the pedestal parameters [2,3]. In such a case, the heating power needed to sustain the core is essentially the heating power needed to sustain the pedestal. This, in turn, is determined by inter-ELM pedestal energy losses, and thus, these determine the denominator of the energy confinement time, $\tau_E$ = Plasma stored energy/Heating power. Pedestal energy transport models of the inter-ELM phase are, therefore, a missing link to a physical and predictive understanding of $\tau_E$. An important step for this is to identify the type of instabilities that are predominantly responsible for energy transport in pedestals of existing experiments- this paper is devoted to answering this very fundamental question.

Although the formation of an ETB is associated with the suppression of instabilities that cause anomalous transport (for instance in an L-mode), there do remain residual instabilities in the pedestal because the observed transport, however reduced, often cannot be wholly explained by neoclassical processes.

In our study of pedestal instabilities (modes), we will follow, here, a novel approach; we will identify the pedestal modes, not through their stability boundaries, but primarily through what they are and what they do. Due to the profound differences in the underlying physical processes that drive different modes, distinct classes of modes have their unique signatures that we will call mode "fingerprints". There are two essential sets of fingerprints that this paper dwells on: 1) the transport fingerprints, epitomized in the ratios between the mode induced diffusivities in various channels, .e.g., the ratio of a particle species diffusivity D to the heat diffusivity $\chi$ of a species 2) the mode frequency in the plasma frame. The principal channels of interest are: the electron and ion temperatures $T_e$, $T_i$, the density $n_e$ and impurity density $n_Z$. With a knowledge of the fingerprints, one can *identify* modes responsible for transport from extant transport observations.

An extremely interesting and rather powerful aspect of this approach is that simple analytical theory (aided by the orderings appropriate to a pedestal: short gradient scale lengths, for instance) can tell us a great deal about the mode fingerprints (being straightforward consequence of the basic physics defining the mode). While fully exploiting analytical and semi-analytical methods, we will also carry out detailed state of the art pedestal plasma simulations to "calculate" mode fingerprints. Insights from the analytical theory and quantitative calculations from simulations will be merged together to paint a far more comprehensive picture of pedestal transport than heretofore.

In a crucial sense, the fingerprints constitute a more robust and useful description of a mode type than its "degree" of linear instability (measured, for instance, by the growth rate); the mode fingerprint is independent of the degree of instability.

The fingerprint methodology turns out to be quite potent- together with the experimentally inferred transport behavior in multi channels, it often severely constrains the choices for modes consistent with observations. The MHD-like modes [5-15], including Kinetic Ballooning Modes (KBM), for instance, can generally be ruled out for energy transport. In fact, for a given transport data the consistent possibilities can often be reduced to only one or two, hugely facilitating effective identification of the mode(s) controlling the pedestal characteristics.

We find that in most existing pedestals, the dominant energy loss culprits are the Electron Temperature Gradient (ETG, [12,15-20] and/or Micro Tearing Modes (MTM, 11,12,15,18,21,22]. In some JET pedestals, major energy losses might result from coupled Ion Temperature Gradient/Trapped Electron Modes.
The MHD-like modes, though not major players in heat loss, might still dominate particle transport.

Application of the fingerprint concept helps us understand several examples of experimentally inferred transport characteristics from the literature: 1) analysis of observations showing that pedestal ion heat transport is neoclassical [23-25], 2) likewise, pedestal impurity transport is neoclassical [26,27] 3) the particle source and effective particle diffusivity in a pedestal is often estimated to be lower than the thermal diffusivity [25,28-32], and 4) the surprising fact that Resonant Magnetic Perturbations (RMPs) typically cause density pump-out rather than flatten the electron temperature in the steep gradient region [33-36]. The consistent and reinforcing pattern formed by all these disparate results is that MHD-like modes are not the dominant energy loss agents; MTM and ETG, rather, are (and possibly ITG/TEM on JET).

Having identified MTM as one likely candidate accounting for energy loss, we also find that it has revealed its presence in measured magnetic fluctuations. Quasi-Coherent fluctuations, such as the JET "washboard modes" [37] that are regularly observed in ELMy H-modes, and similar phenomenon on DIII-D [6] and ASDEX-U [38], have multiple signatures of MTM, and cannot apparently be easily explained in terms of other instabilities.

Certainly there are other mode characteristics that can be used to identify fluctuations. For example, when two different fluctuating quantities can be measured, their ratio can distinguish between instabilities. Such techniques have been employed successfully in the past in pedestals [39,40] and the core [41]. In the case of the MAST pedestal, the ratio of $(\delta B/B)/(\delta n/n)$ for pedestal fluctuations indicated that ETG modes were operative during the $T_e$ profile evolution [39]. A different ratio $(\delta T/T)/(\delta n/n)$ was considered on TFTR [41]. However our emphasis here is, primarily, on ratios of the *transport* caused by the fluctuation in different channels. Not only can these ratios be used to *identify* the modes, but as we will see, they can also *determine the role* of an instability in pedestal evolution. One of our conclusions here is that these ratios, for pedestal parameters, are closely connected to the fluctuation frequency in the plasma frame. Hence this is another fingerprint we will consider. However the basic framework presented here can be generalized to include other possible fingerprints. For example, the same analytic arguments that are used to obtain the transport fingerprints of MHD-like modes and of MTM can also be used to indicate that ratios of two different fluctuating quantities (e.g., $(\delta B/B)/(\delta n/n)$, $(\delta T_e/T_e)/(\delta n/n)$ and $(\delta B/B)/(\delta T_e/T_e)$ ) should also be quite different for these two modes as well.

One of the principal tasks of this paper was to thoroughly "fingerprint" two well diagnosed DIII-D pedestals. For one of these (shots #153674/153675), we infer the pedestal transport from the reported inter-ELM evolution of pedestal profiles of the electron and ion temperature $T_e$, and $T_i$, and electron and impurity density $n_e$ and $n_Z$ [5]. Fortunately for the second (shot 98889), a detailed multi-channel transport analysis was available in the literature [24].

Painstaking gyrokinetic simulations of both cases were performed. Important conclusions include that 1) the fingerprint concepts were thoroughly corroborated numerically 2) nonlinear simulations find that quasi-coherent MTM, with quasi-coherent magnetic spectra qualitatively similar to those that are widely observed, are capable of producing much, or all, of the transport power, 3) the simulated MTM, in addition to giving substantial energy losses, can simultaneously have frequencies in the lab frame that are a reasonable match to the actual fluctuations measured on each shot 4) MHD-like modes, MTM and/or ETG modes constitute the instabilities that are likely to survive in a pedestal (less prone to shear suppression). Identification to distinguish between MHD like modes and MTM/ETG is possible based on the fingerprints mentioned above and simulations guided by them.

It is important to emphasize here that the above results do not contradict the EPED model [3,4]. MHD-like modes such as KBM might still cause the dominant density transport without much contribution to the energy transport. MHD-like modes may control the density profile to impose marginal stability on inter-ELM pressure profiles, consistent with assumptions of the EPED model.

With this motivation and introduction, we outline the scope of this paper.

In section II we apply the fingerprint concept to experimental observations of transport in various channels, to identify modes responsible for transport. In section III we describe why the fingerprint approach is often crucial in order to identify pedestal transport, relative to the conventional approach of examining pedestal stability of particular cases. In section IV, we show how the fingerprints result from the underlying physics of the modes using easily accessible physical arguments. In section V we apply the fingerprint concept, in detail, to two DIII-D cases, including gyrokinetic calculations for those pedestals. Both fluctuation characteristics and transport observations are considered. In section VI, we give an analytical description to magnetic modes, based on the drift-kinetic equation, and derive the fingerprints for a pedestal gradient ordering. These are compared to gyrokinetic results using GENE. Finally, in section VII further details of the application of GENE to the DIII-D discharges are presented. We summarize and indicate future applications in section VIII.

## II Application of the fingerprint concept to experimental observations

First, we start with some definitions. The transport diffusivity for each channel and species is, conventionally, defined via flux divided by the gradient. The particle and heat diffusivity $D_s$ and $\chi_s$, respectively, for instance, are

$D_s = \Gamma_s/(dn_s/dx)$ 					Eq(1)

$\chi_s = [Q_s - (3/2) T_s \Gamma_s ]/(n_s dT_s/dx)$ 			Eq(2)

Where s denotes the species.

Next, we summarize the transport fingerprints of relevant pedestal modes. The Gyrokinetic simulations of current experimental pedestals found that a variety of unstable modes could cause transport in some mode-specific channels (including in our detailed results for several DIII-D cases). We have broken up the modes (and their effects) into two categories [tables 1A and 1B] ; the first set is resistant to velocity shear suppression, while the second set can be shear suppressed.

TABLE 1A Approximate fingerprints for of modes resistant to shear suppression

| Mode type | $\chi_i/\chi_e$ | $D_e/\chi_e$ | $D_Z/\chi_e$ |
|---|---|---|---|
| MHD-like | 1 | 2/3 | 2/3 |
| MTM | ~1/10 | ~1/10 | ~1/10 |
| ETG | ~1/10 | ~1/20 | ~1/20 |

TABLE 1B Approximate fingerprints for modes susceptible to shear suppression

| Mode type | $\chi_e/\chi_i$ | $D_e/(\chi_i+\chi_e)$ | $D_Z/(\chi_i+\chi_e)$ |
|---|---|---|---|
| ITG/TEM | 1/4 - 1 | -1/10 – +1/3 | ~1 |

It is from these candidates that we will extract possible culprits for pedestal transport.

The results, displayed in the tables, are not qualitatively surprising, but they embody, as we will show, a deep message. Therefore, we have invested considerable effort in confirming these results, both analytically, and via simulations. These calculations are described in sections IV and VI, and they establish these ratios with considerable generality and detail. They focus on the unique pedestal parameter regime for the first time, which differs from the core in important ways.

The MHD-like modes (including KBM) cause very comparable diffusivities in <u>all</u> channels, and <u>no</u> pinches. The MTM and ETG, on the other hand, cause almost exclusively electron thermal transport. ITG/TEM are the most complex, but nonetheless, they can be characterized as causing thermal transport in both electrons and ions, and they cause strong impurity transport. Depending on parameters, the associated electron particle transport can range from moderate to a weak diffusion accompanied by a weak pinch ($D_e/(\chi_i+\chi_e)$ is slightly negative). Even accounting for uncertainties for ITG/TEM, the above results are consequential.

The three modes enumerated in Table 1A- MHD-like, ETG and MTM- are resistant to velocity shear suppression. Consequently, these are expected to be the primary candidates for anomalous transport in most current pedestals that tend to be highly sheared. Earlier simulations in the literature [8,15-20], and analysis and simulations here, strongly support this expected behavior. Since the modes listed in Table1B may be excited in pedestals of some larger machines of today, and of future (lower velocity shear) devices [18-20,42], we will also include ITG/TEM modes [15-20,42-45] in our investigations, both for completeness and for future applications. It must be, however, stressed that these modes are generally suppressed for many pedestal parameters on present tokamaks; in fact, their suppression is what leads to the very formation of the pedestal.

As a summary, we have included a block diagram of how we apply the fingerprint concept.

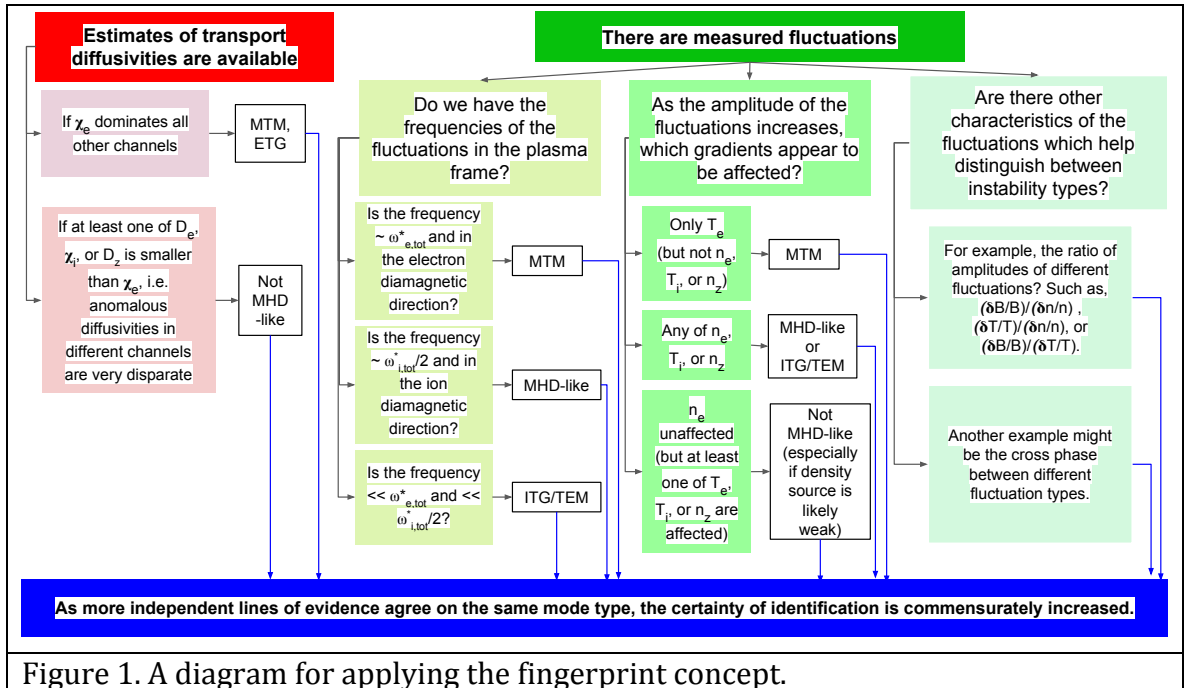

Figure 1. A diagram for applying the fingerprint concept.

In this paper we emphasize the fingerprint aspects related to the transport, and also the frequency. However, the fingerprint concept is evolving, so we include the possibility for using other potential fingerprints as well (though the details of such extensions will be left to future work).

Ion heat transport is sometimes observed to be neoclassical

On a substantial dataset of ASDEX-U shots that were analyzed in detail [23,24], the ion thermal diffusivity $\chi_i$ is always in rather good agreement with the predictions of neoclassical theory, $\chi_{neo}$. For *some* of these shots, the electron transport is highly anomalous, with $\chi_e \sim \chi_i$. A detailed multi-channel transport analysis of a DIII-D shot comes to similar conclusion [25]: $\chi_e$ is anomalous and greater than $\chi_i$, while the latter is neoclassical. There are, thus, two distinct sub categories: 1) $\chi_i \sim \chi_{neo}$ and substantial anomalous $\chi_e$, $\chi_e \geq \chi_I$, and 2) $\chi_i \sim \chi_{neo}$ but low electron transport, $\chi_e \ll \chi_i$ (observed on ASDEX-U [23]).

The electron and ion transport, characteristic to sub-category 1, could not be driven by MHD-like modes, since their transport fingerprints demand a comparable anomaly in the ion and electron channels, that is, we would have $\chi_e \cong \chi_i - \chi_{neo}$. Instead, we have $\chi_e \approx \chi_i$, and $\chi_i \sim \chi_{neo}$; the source of anomalous $\chi_e$, therefore, have to be some other class of instabilities. The most relevant class, in addition, has to be resistant to shear suppression (so that they can cause significant transport). Hence, among the selection of modes that are most likely significant in pedestals (MHD-like, MTM, ETG, and ITG/TEM), some combination of MTM and ETG must be responsible for anomalous energy transport- consistent with neoclassical $\chi_i$ simultaneously with a substantial anomalous $\chi_e$.

Ion scale electrostatic modes are likely to be suppressed due to velocity shear for ASDEX-U parameters according to gyrokinetic simulations [19-20]. Based on the experimental observations just noted, such modes, indeed, are not a major energy loss agent because no large anomaly in the ion channel was observed.

For the second subcategory ($\chi_e \ll \chi_{neo} \approx \chi_i$), ion neoclassical transport is apparently the dominant energy loss channel; evidently, energy transport induced by MHD-like or any other instability, is of relatively minor importance.

Impurity transport is similar to neoclassical

The impurity transport channel might be an even more sensitive indicator of turbulent transport than is the ion thermal transport. First, we briefly discuss the expected impurity behavior in neoclassical theory.

Neoclassical theory predicts an impurity flux $\Gamma_Z$ with diffusion $D_Z$ and a pinch $V_Z$:

$$\Gamma_Z = D_Z\, dn_Z/dx - n_Z\, V_Z \qquad \text{Eq(3)}$$

where $n_Z$ is the impurity density. Since there is very little impurity source in the pedestal, in steady state $\Gamma_Z=0$, and the solution for $n_Z$ can be easily obtained analytically. The ratio of impurity densities at two locations a and b is

$$n_Z(a)/n_Z(b) = \exp\left(\int_a^b dx\, V_Z/D_Z\right) \qquad \text{Eq(4)}$$

Note that this is exponentially sensitive to $1/D_Z$. The neoclassical pinch $V_Z$ is $\sim Z$, so this sensitivity increases with Z. Thus, neoclassical theory predicts a strong, Z dependent, enrichment of impurities at the top of the pedestal in steady state. This is, in fact, observed experimentally [26,27].

Due to the exponential sensitivity, a turbulent $D_Z$ would produce a strong modification from the neoclassical value for $n_Z(a)/n_Z(b)$. Because of this expected strong effect, published analysis on ASDEX–U were able to conclude that "turbulent transport was of negligible importance to impurity transport" [26].

With this preamble, let us discuss commonplace experimental observations regarding impurities in H-modes.

Without ELMs, high Z impurities tend to build up, and core radiation rises to high levels that often terminate the H-mode (unless some atypical fluctuation is present, as in QH modes [46] and I-modes [47]). For typical ELMy H-modes, the impurity density typically rises in the inter-ELM phase, and ELMs are needed to expel them.

The ELMs are, evidently, far more effective at expelling impurities (particularly with high Z) than is the typical inter-ELM transport. This is rather surprising, since ELMs expel only ~ 20-40% of the net heating (on time average [48]), and the remainder of the energy is expelled by the inter-ELM energy transport. In other words, inter-ELM transport is much more effective at expelling energy than impurities, in comparison to the ELM MHD instability.

Let us examine these observations in the light of the fingerprint concept:

> A) If an inter-ELM MHD-like mode (e.g. KBM) were responsible for most inter-ELM energy losses (i.e. $\chi$), a largely enhanced impurity diffusivity $D_z \sim \chi$ will lead to a strong expulsion of impurities; the impurity build-up from the neoclassical pinch will, then, be strongly reduced or eliminated. If this were so, then why would an MHD instability in the form of an ELM be critically necessary to expel high Z impurities? This contradiction is even more acute since the ELM instability expels less energy than transport between ELMs- why would an MHD instability in the form of an ELM be needed if an MHD-like instability were present in the inter-ELM phase which was an even stronger transport agent?
>
> B) But if inter-ELM turbulent energy transport is dominated by MTM and ETG, *which have little effect on impurities*, then the need for ELMs to expel impurities is clear.

The observations of pedestal impurity transport reinforce all the same conclusions reached by analyzing ion neoclassical thermal transport. Specifically: 1) experimental behavior is not consistent with MHD-like modes being the dominant heat transport mechanism in typical ELMy H-mode pedestals. It is also not consistent with ITG/TEM modes being the dominant energy loss. But it is consistent with MTM and/or ETG playing that role.

## The electron particle source is low

This source in the pedestal has been estimated on various devices [25,28-32]. Here we present results for several JET shots. The electron particle source was found to be small in the relevant sense.

The source is dominated by neutral penetration and ionization in the pedestal. Edge modeling codes use edge data of various kinds to constrain the simulation.

The density source S can be used to compute the inferred $D_e$, $D_{inf}$, from $S = D_{inf}\, dn/dx$: $D_{inf}$ is the inferred diffusion needed to exhaust the source. Similarly power balance can be used to compute $\chi_{inf} \approx \chi_i + \chi_e$, to arrive at the fingerprint $D_e/(\chi_i + \chi_e)$. We use the sum of $\chi$ for the following reason. While the total power flow through the pedestal can be estimated with reasonable accuracy, the fraction going, separately, through the ion channel and the electron channel is much more uncertain. In such cases, the sum $\chi_i + \chi_e$ has considerably less uncertainty than each term individually. Hence, the fingerprint $D_e/(\chi_i + \chi_e)$ can be inferred from the data with less uncertainty, and compared to the same fingerprint for specific modes.

The JINTRAC code was used to estimate the source for shot 84794 (as described in reference [31]. The value of $D_e/(\chi_i + \chi_e)$ in the pedestal was $\sim 0.07$. This is far smaller than the value expected for an MHD-like mode, which is $1/(3/2 + 3/2) = 1/3$. The disparity rules out MHD-like modes as primary agents for energy transport.

This shot had $I_p/B = 1.4$ MA/1.7T. Nonlinear simulations of other profiles for such conditions find that ITG/TEM modes are likely suppressed as a dominant total energy loss process. However, profile specific effects might be significant so they cannot be completely ruled out for this shot. It is safer to conclude that energy losses are dominated by some combination of MTM, ETG, or, possibly, ITG/TEM, to the extent the latter can escape suppression.

The edge code EDGE2D [32] was used to estimate the sources in JET shots 92168 and 92174. The estimated values of $D_e/(\chi_i + \chi_e) \sim 0.07$-$0.1$, are small (as in 84794 above), fortifying the conclusion that MHD-like modes do not control energy losses.

In fact, independently of the results of edge codes, the fingerprint concept together with experimental observations can imply that the density source must be low, in cases considered above: where neoclassical transport pertains to the ion thermal channel, or, the impurity channel.

Let us now argue why the density source is small in these cases.

1) If the density source was substantial so that $D_{in}/\chi_{inf} \sim 1$, we will need MHD-like / ITG modes for exhausting particles. Such modes, if called upon to produce a requisite $D_e$, will cause a substantial anomaly in the ion thermal channel- which is not found. In addition, a large anomalous impurity diffusivity would arise, which is also inconsistent with experiment.
2) On the other hand, if MTM and/or ETG were called upon to exhaust a large particle source, a large energy loss, inconsistent with power balance, would automatically follow.

Both of these scenarios are inconsistent with observations; the only consistent possibility is that the density source must be small.

And as we see below, the smallness of the density source has a special importance.

Occam's Razor

We have just concluded, using three different pathways, that MHD-like modes do not dominate the energy transport. How do we square this with the recent history of MHD-based models being able to "predict" (reasonably accurately) the pedestal pressure? Models such as EPED [4,5] posit that MHD-like modes (KBM) impose marginal stability in the inter-ELM phase. Widespread phenomenological observations also seem to indicate that MHD-like modes may be enforcing marginal stability before an ELM (see section III). The question naturally arises, then, how this can be consistent with energy transport being dominated by MTM and/or ETG modes.

The resolution of this is that MHD-like modes can enforce marginal stability by dominating the density transport, when the density source is small.

In fact, in the spirit of Occam's razor, the transport fingerprint of the relevant modes with the ansatz of a weak density source, gives the simplest explanation for the disparate experimental facts (as well as the RMP observations to follow).

Consider an idealized thought experiment. As the pressure gradients increase following an ELM, an MHD-like mode (e.g. KBM) is excited, causing comparable diffusivity in all channels. Since the $n_e$ source is weak, it is much more strongly affected- so its evolution, mainly, enforces MHD marginal stability (as in EPED [3,4]). *Since the density source is weak, a small $D_e$ will suffice to strongly affect the density profile. But the $\chi_e$ and $\chi_i$, needed to saturate the evolution of $T_e$ and $T_i$ are much larger than $D_e$, since those channels are much more strongly driven. The requisite $\chi_e$ and $\chi_i$, therefore, must arise from MTM, ETG and neoclassical.*

*If MHD-like modes did cause most energy transport, $n_e$ would quickly collapse, stabilizing them.*

The weak density source also ensures that the impurity transport will not be strongly affected: the smallness of $D_e$ also implies that the impurity diffusivity $D_Z$ will be small for MHD-like modes. And MTM and ETG also give low $D_Z$.

The thought experiment shows that the role of an instability depends, crucially, upon factors that, till now, have gone largely unappreciated:

1) *Relative effects of the instability on different transport channels*
2) Relative strength of sources, e.g., *a weak $n_e$ source*
3) Marginal stability characteristics of the mode

The *three characteristics enumerated above, can, together, prevent a strong effect on T from MHD-like modes even when $n_e$ is strongly affected.*

Resonant Magnetic Perturbations (RMPs)

Surprisingly, experiments with externally imposed resonant magnetic perturbations, RMPs, observe induced transport [33-36] that provides further confirmation of the conclusions in the sections above. This is because (as we will see later) the RMPs, when the self-consistent plasma response is included, can be viewed as externally seeded MHD-like instabilities in the steep gradient region. Though static in the lab frame, the large ExB Doppler shift (pedestal have large $E_r$ ) imparts them, *in the plasma frame,* an effective frequency $\omega \sim$ ion diamagnetic $\omega_i^*$, typical of ideal MHD-like modes. Anticipating section VI, this similarity implies the self-consistent plasma response gives $\delta E_{||} \sim 0$ for the RMP in the steep gradient region. Whatever the magnitude of the total magnetic perturbation (including screening), the transport fingerprint of RMPs in the steep gradient region will resemble those of an MHD-like mode.

Viewing RMPs like MHD modes with unique fingerprints greatly facilitates the interpretation of a crucial experimental observation: RMPs are observed to result in density pump-out, reduction in pedestal $n_e$, but have relatively little effect on the pedestal $T_e$ or $T_i$ in the steep gradient region [33-36].

Hence, the observations reinforce the point that MHD-like modes cannot be the dominant energy transport mechanism in pedestals: if such instabilities *could* play this role, then the application of RMP would flatten temperature profiles. But this does *not* happen, rather, density flattens much more strongly. Hence, if MHD-like instabilities arise, then they will enforce marginal stability of the pressure profile in the steep gradient region mainly by reducing the density gradients.

We will derive in section VI, an approximate criterion that the self-consistent electrostatic potential perturbation leads to $\delta E_{||} \sim 0$; for RMP, it reads

$$1 \gg (m_i/m_e)(L_{ped}/L_s)^2 \qquad \text{Eq(5)}$$

For all pedestals we have examined, in the steep gradient region, this is satisfied. (In mid pedestal, $L_{ped}/R \sim 1/100$, $L_s \sim qR$, $q \sim 3$, so $(L_{ped}/L_s)^2 \sim 10^{-5}$, and $(m_i/m_e)(L_{ped}/L_s)^2 \sim 1/20$). Hence, RMPs can be considered as an externally seeded MHD-like fluctuation in this region. (But inside the pedestal top, $L_{ped}$ is an order of magnitude larger, and this does not hold.)

Thus, the observed effects of RMPs can be considered an experimental test of the pedestal transport response to MHD-like instabilities. They are found to cause density pump-out, but do not strongly affect the temperature gradients in the steep gradient region. Hence, they cannot be responsible for most energy transport, but they might dominate electron particle transport.

Summing Up

A remarkably coherent and consistent picture has emerged. Multiple independent and disparate lines of evidence reach a similar conclusion, namely, that some combination of MTM and/or ETG dominate energy transport in typical experimental pedestals. MHD-like modes do not dominate energy transport, but rather, might dominate electron particle transport (and potentially enforce MHD marginal stability by that mechanism).

This conclusion rests on very diverse types of observations, and the consistent pattern they coalesce toward, when clarified through the lens of the transport fingerprint. The cumulative effect is significantly stronger than any single line of evidence.

Or from another point of view, the diverse experimental observations might be considered as having a unified explanation, that is, they may be simply explained as arising from a single ansatz, namely a weak density source, together with the theoretically inferred transport fingerprints of the modes widely expected to be responsible for transport.

**III The imperative for the transport fingerprint concept to augment conventional approaches**

In order to appreciate the central role of the fingerprint concept in advancing the understanding the pedestal, we must critically examine what is, perhaps, the most common approach to determining the pedestal transport: one takes experimentally reconstructed equilibria and feeds them into a code to determine the instabilities and nonlinear transport arising for that profile. Despite the appeal of the straightforwardness of this approach, it is unlikely to unravel the pedestal transport problem, because of the presence of multiple instabilities close to thresholds, together with significant uncertainties in the data and models used for analysis. The transport fingerprint concept, allows progress, because it "measures" a mode not by its instability characteristic, but by several other defining features. Let us see how and why.

Consider MHD-like modes first. Typically, pedestal pressure profiles come to a quasi-steady state in the inter-ELM phase, which is maintained for a substantial fraction of the ELM cycle before an ELM occurs. The fact that the ELM occurs is strong prima facie evidence that the pressure profile is very near the MHD stability boundary in this phase. And MHD stability calculations of peeling ballooning modes usually find that the later stages of the inter-ELM phase are at the stability boundary, within the error bars.

Since the profile is close to this boundary for a considerable time before the ELM, it is very reasonable to posit that an MHD-like mode might be operative in this phase, causing transport.

MHD-like modes are very strong, especially in comparison to the weak transport in the edge barrier, and their transport should be very stiff. Rough estimates based on linear calculations here (estimating the diffusivity by the mixing length estimate $\gamma/k_\perp^2$) find that transport at the level of inter-ELM power balance might arise if the stability boundary is exceeded by ~ one percent. This is qualitatively consistent with nonlinear gyrokinetic simulations [8], which find that there are very small differences between profiles that are stable, and, ones that give transport one and a half orders of magnitude higher than power balance.

It is highly unlikely that either reconstructed experimental equilibria, or, the theoretical boundary for onset of MHD-like transport matching power balance, will be known to such high accuracy in the foreseeable future. *So with realistic errors, applying the conventional methodology, a code applied to a reconstructed profile would likely calculate MHD transport to be anywhere between insignificant to overwhelming. By varying within the error bars, effectively any transport level could be attained for any particular channel.*

Such uncertainties are by no means limited to MHD-like modes. Specifically, ETG modes are often rather close to an effective nonlinear transport threshold, where significant transport arises compared to the heating power. This is true for one of the DIII-D shots considered here, and other simulated cases in the literature [15-20]. For these cases, transport changes by a factor of ~ 2-3 for a ~10-20% change in the profile gradients.

For MTM modes, we find in section V and VII that there is also a surprising sensitivity to the profiles for the linear mode spectrum, growth rates, and transport.

Specifically, we have examined transport sensitivity for a DIII-D shot that we have simulated in detail (shot 153674/5, presented in sections V and VII). Self-consistent equilibria for several profiles, that are apparently within the error bars for shot 153674/5 (see figure 2), were studied. The simulated ETG transport varies by nearly an order of magnitude for these similar profiles- from insignificant to excessive. (See sections V and VII.) 2) The MTM nonlinear behavior and heat fluxes are strongly sensitive to the profiles, varying by several fold (see section VII).

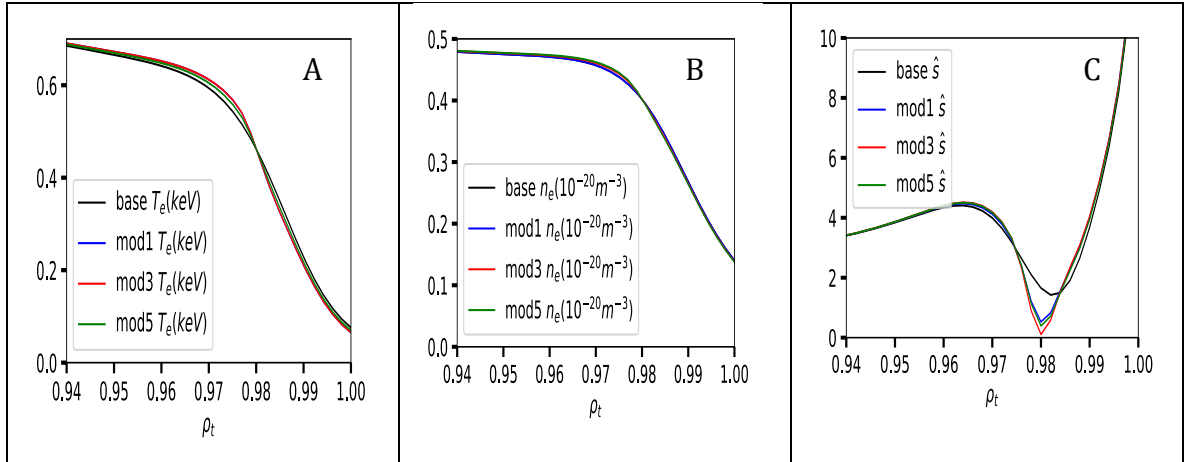

Fig 2 In (A), we show the tanh fit $T_e$ profile (the base case), and the three modified profiles called mod1, mod3 and mod5. In (B), the density profiles are shown. The magnetic shear is strongly affected, $\hat{s} = (\rho_t/q)dq/d\rho_t$, where $\rho_t$ is the radial coordinate (normalized toroidal flux)$^{1/2}$.

Uncertainties arise not only from the experimental profile reconstruction, but also, from computational model incompleteness. Many experiments observe Quasi-Coherent Fluctuations that are associated with $T_e$ transport (e.g. JET washboard modes [37], observations on DIII-D [6], ASDEX-U [38], and Alcator C-mod [13,14]). Our simulations of such modes show, as one would expect to be true universally, that some localized flattening of the temperature profiles is an important aspect of nonlinear saturation for quasi-coherent modes. In view of the likely importance of this mechanism, it should be mentioned that simulations do not include physically important aspects like:

    1) By flattening the temperature profile in the region where the instability is driven, the bootstrap current in the magnetic equilibrium is changed, which dynamically changes the q profile and magnetic shear during the evolution- and this is important for stability. No gyrokinetic code presently includes this, to our knowledge.

    2) The modified $T_e$ profile from a QCF also changes ETG transport in the region affected. So, a correct description of the $T_e$ profile evolution from QCF should simultaneously include the modified ETG transport, which acts to

further modify the $T_e$ profile, further altering QCF saturation. A single such multi-scale simulation would expend several tens of millions of CPU hours (with GENE, and much more for some other codes), which is extraordinarily expensive, and has never been attempted (much less with the requisite sensitivity studies for multiple profiles within the error bars).
3) In addition to these specific physical effects, the gyrokinetic simulation of strongly electromagnetic instabilities in pedestals is very numerically challenging in other ways. Some of these aspects are described in section VII, and, at present, require the use of approximations that could affect quantitative accuracy.

The message from these considerations is that computed transport from MHD-like modes, ETG, and MTM is significantly uncertain. For completeness, we feel we should briefly mention some additional question marks. These include multiple effects that are not included in the usual error bars quoted for pedestal reconstructions 1) the appropriateness and fidelity of the simple fitting functions used to very considerably smooth the raw pedestal profile data 2) measurement uncertainty in the relative alignment of electron and ion quantities measured by different diagnostics 3) various uncomputed physical effects in the calculation of the bootstrap current, especially 4) order of magnitude variations of impurity density (poloidally) within a flux surface, that are expected and measured, and 5) the large poloidal gyroradius of bulk ion and beam species. All these effects are likely more significant for modes that are somewhat localized in a pedestal, such as ETG and MTM, than for more global MHD-like modes which average over the pedestal. Uncertainties in the local q profile and magnetic shear affect the local proximity to thresholds for ETG and MTM, which in turn effects stability and transport.

In view of all these uncertainties, one would be very unlikely to unravel whether most energy losses were from MHD-like modes, as opposed to MTM and/or ETG, by conventional approaches. By varying profiles and other approximations in one way or another within the error bars, any mode might dominate energy losses, or, transport in other channels.

Amazingly, the injection of the fingerprint concept into computational and analytic theory works wonders. Despite all the uncertainties about *how unstable* a mode is, by changing the focus away from degree of instability, and toward the *unique transport signatures of the viable candidates should they be present* (and also, fluctuation frequencies in the plasma frame, if available), identification of the controlling instabilities becomes possible.

**IV. Physics of the mode types and their transport fingerprints**

We now turn to consider how the fundamental physics of pedestal instabilities is reflected in their transport fingerprints. Here we recapture some salient features of the relevant instabilities. In section VI, a more detailed analytical description of the fingerprints for MHD-like modes and MTM will be given.

MHD-like modes

Given the fact that ideal MHD is a *one*-fluid model, different species (contributing to this fluid) *inevitably behave the same way*. Transport diffusivities in all channels are, then, expected to be the same. Furthermore, the good correlation between ideal MHD stability calculations and experimental ELM thresholds indicates that the actual pressure response of the mode is not very far from the ideal MHD prescription.

Energetically, any instability grows by tapping the free energy in the equilibrium. In confined plasmas, this energy is supplied by reducing the equilibrium gradients. MHD modes are driven by pressure gradients and current gradients, and gradients from either temperature or density give about the same contribution to the growth rate, and hence, give about the same rate at which the equilibrium pressure gradient is reduced. This is also independent of species. So, using the definition of diffusivity in section 2, ideal MHD modes produce comparable diffusivity in all channels and all species.

Arguments based on the dynamical equations describing the pressure and the density can also be used to come to this conclusion. In pedestals, the density and pressure evolution are dominated by perturbed ExB convection. Effects from compressibility can lead to modifications of the pressure response from this purely convective response. But pedestals have two small parameters that greatly simplify the equations so that ExB convection dominates: 1) the perpendicular scale L is enormously less than R, the major radius, $L/R \sim 10^{-2}$; and the latter is the scale where perpendicular compression is important 2) The sound speed $c_s$ is much less than the Alfven speed, $(c_s/v_A)^2 \sim \beta$, so that parallel compression is not large.

Using these small parameters, the dynamical equations for ideal MHD for perturbations around a static equilibrium becomes $d\, \delta p/dt = \mathbf{v}_E \cdot \nabla p_0$ (and similarly for density), where subscripts 0 denote equilibrium quantities. Some details are given in Appendix 1. When all plasma perturbed quantities are almost purely convective with the same ExB velocity, it is to be expected that gradients of density and temperature for all species will relax at about the same rate, i.e., for MHD modes, the transport diffusivities in all channels: $T_i$, $T_e$, $n_e$, and impurity density $n_{Z:}$ will be very similar.

It is generally appreciated that MHD is a surprisingly rugged description of plasma behavior, and is often valid outside of the nominal assumptions that are used in its derivation. Let us consider how this pertains, specifically, to the transport fingerprints of pedestal modes.

Because of the steep gradients, the diamagnetic frequency ($\omega^*$) in the pedestal, is much larger than in the core; it can easily be of order of, or even larger than parallel Alfvenic frequencies $\omega^* \sim \omega_A = v_A/qR$ (where $v_A$ is the Alfven speed, q the safety factor, and R the major radius). However, diamagnetic effects can be included in reduced *one fluid theories* (i.e., Ref [49], which also includes compressibility), and these have all the same properties noted above for ideal MHD.

We extend such arguments to a kinetic analysis here. As we will see in our fully kinetic treatment of fluctuations in section VI, the condition $\delta E_{||} \sim 0$, together with the smallness of L/R and $\beta$, suffices to give comparable quasi-linear transport in all channels, when kinetic effects, e.g., trapped particles, resonances, etc., are included. Importantly, in addition, there are no pinches for MHD-like modes; transport is diffusive.

The magnitude of linear growth rates of MHD-like modes well past threshold can be roughly estimated analytically, and they greatly exceed the velocity shearing rate in a pedestal. This has, essentially, already been indicated in the EPED model literature [4,5] for the case of KBM, and it is part of the justification of the EPED ansatz that KBM enforce marginal stability in the inter-ELM phase. We also corroborate that growth rates can easily exceed shearing rates for the DIII-D cases simulated here. Hence, MHD-like modes are, generally, candidates to cause pedestal transport.

Finally, we note that when MHD modes are close to marginal stability, various kinetic effects can modify their threshold. However, these modes are important precisely because the growth rate of primarily MHD modes can be so high that they can overcome velocity shear suppression. When they are so unstable, their dynamics are primarily MHD-like. Analytical arguments given below indicate that $\delta E_{||} \sim 0$ for such modes in a pedestal, kinetic effects do not substantially change their fingerprint, and also, gyrokinetic simulations for two DIII-D pedestals corroborate this.

Micro Tearing Modes

The arguments in section II find that MTM are particularly important candidates to cause pedestal energy losses. These modes also produce magnetic fluctuations that might be readily observable. So we consider MTM here in somewhat greater detail than the other candidates. It is relevant to ask whether there is any other supporting evidence for their presence, that is, in addition to the calculations for specific shots that show that they can be unstable. Certainly, more detailed gyrokinetic calculations of multiple profiles and devices are needed. However, we might also ask if the characteristics of these modes is such that it is likely that they are often unstable, and also, if some observed magnetic fluctuations have qualities that make it likely that they are MTM rather than MHD-like modes. In this section, we answer both of the latter questions affirmatively. The transport fingerprints of MTM are an important element of this; they are particularly simple, and follow from their underlying physics.

MTM are a particular example of Resistive MHD modes (RMHD modes), which differ from ideal MHD-like instabilities (which we usually refer to as simply MHD-like modes) in that finite resistivity is essential. The RMHD modes and ideal MHD-like modes are the two major candidates to explain observed magnetic signals from the pedestal. Drawing upon the extensive analytic literature on RMHD instabilities [50-56] (and others too numerous to list), it becomes apparent that the unique parameter regime of a pedestal strongly favors the Micro Tearing Mode variant of such modes. And indeed, gyrokinetic simulations of pedestals find MTM (references [10,11,15,18,21-22] and also our DIII-D simulations in section VI and VIII). We now elucidate the fundamental physical reasons for this.

Most RMHD modes considered in the literature have exactly the same driving mechanisms as ideal MHD, but the presence of resistivity allows reconnection near a rational surface, which enables instability well below the ideal stability boundary. Such modes are obtained in conventional fluid treatments, so we will refer to them as fluid RMHD modes. The instability drive of MTM is totally different from this. As described in the seminal papers [51,52], *kinetic* effects lead to instability when the mode frequency $\omega$ is roughly in the range of the electron collision frequency $\nu_e$, $\omega \sim \nu_e$. (The velocity dependence of the collision frequency is a critical ingredient for these MTM [51,52]. Additional collision-less kinetic destabilization mechanisms have also been found [21,57]. )

The differences between fluid RMHD modes and MTM lead the latter to dominate in pedestals. Because the fluid RMHD are enabled by reconnection, they rely crucially on the relatively slow process of resistive magnetic diffusion near rational surfaces. This is reflected in their growth rates, which depend somewhat on this slow rate, so they grow much more slowly than ideal MHD modes (which proceed at an Alfvenic rate $\omega_A$). For a given level of MHD driving, fluid RMHD instabilities attain their largest growth rates when $\omega^*=0$ (which we refer to as $\gamma_{0Fluid}$). But pedestal levels of $\omega^*$ are very large, and can easily be order $\omega_A$, which far exceeds $\gamma_{0Fluid}$. Extensive literature shows that diamagnetic stabilization of fluid RMHD modes is strong when $\gamma_{0Fluid} \ll \omega^*$ [50-56]. The properties of MTM are the opposite in almost every way, and analytic treatments [51,52] show that 1) their growth rates, $\gamma_{MTM}$, scale as a roughly constant fraction of $\omega^*$, and so, become *larger* as $\omega^*$ increases  2) $\gamma_{MTM}$ is not tied to other rates that lead to stabilization of fluid RMHD modes at large $\omega^*$, in particular, to the slow magnetic diffusion rate. The lowest order MTM dispersion relation for large $\omega^*$, from the first paper to describe the MTM driving mechanism [51], is shown in figure 3A. The kinetic destabilization mechanism is strongest in the vicinity of $\nu_e/\omega^* \sim 1$, but is still operative over a wide range: destabilization at about a third of the maximum value arises over a range of $\nu_e/\omega^*$ of almost two orders of magnitude, of about 0.06 to 3.

(Details for the analytic expressions used in this section are given in the Appendix: Analytic expressions for MTM and fluid Resistive MHD modes.)

The trends with $\omega^*$ mentioned above are also displayed in figure 3B and 3C, for two typical JET cases where magnetic fluctuations, termed "washboard modes", were observed; the modes were inferred to originate from the pedestal region [37]. We use analytical dispersion relations to compare the MTM, and also, the most unstable fluid RMHD modes (with the maximum MHD driving, just before ideal MHD instability arises; see the Appendix for details). As seen in Fig. 3B and 3C, fluid RMHD modes are significant only at unrealistically small values of $\omega^*$. The MTM are overwhelmingly dominant for the large levels of $\omega^*$ typical in pedestals.

In Table 2, we display rough estimates of MTM and fluid RMHD modes, for multiple pedestals on DIII-D, JET and ASDEX-U where magnetic spectrograms or other diagnostics showed fluctuations with a particular toroidal mode number n [6,37,38]. To allow estimates, we presume the modes originate from the middle of the pedestal, and estimate parameters there from the pedestal top values. (See the Appendix for details). As can be seen, for all cases, $\nu_e/\omega^*$ is in the range of substantial MTM drive. Also, the pedestals are in the regime where fluid RMHD modes are strongly stabilized by diamagnetic effects. Even allowing for the roughness of these estimates, it is clear that MTM are the dominant resistive MHD modes.

We stress that it is very important that the references which characterize the instances of fluctuations in Table 2 (of washboard modes on JET [37], the high frequency fluctuations on ASDEX-U [38], and DIIID shot 153674 [6]), describe them as having both 1) a close association with electron temperature profile evolution and electron energy transport, and 2) the growth of the fluctuations has no apparent effect on electron particle transport.

Presumably, a growing fluctuation amplitude is causing a growing transport. The fact that during this phase of growing amplitude, the density profile is unaffected, but, the temperature gradient is ultimately saturated, is inconsistent with MHD-like modes. This is especially true in view of the relatively small density sources in a pedestal (as estimated in multiple devices, see section II): MHD-like modes would be expected to affect the density channel even *more strongly* than the $T_e$ channel.

The observed transport fingerprint of these fluctuations is fully consistent with these modes being MTM. Furthermore, the analytical arguments indicate that MTM are likely to be unstable in pedestals, whereas other RMHD modes are not. While further detailed analysis on specific cases is clearly necessary, this evidence, taken together, indicates that these magnetic fluctuations are likely MTM.

Published descriptions for the DIII-D and ASDEX-U shots have noted that the fluctuations observed seem quite similar to the JET washboard modes. Detailed analysis of two DIII-D shots here show that their observed magnetic fluctuations are indeed MTM. The general results here indicate that the similarity inferred in the experimental papers is likely physically well grounded: the modes are probably the same basic type of instability, namely, MTMs.

Table 2

| Discharge | Toroidal mode n | $\gamma_{MTM}/\omega^*$ | $\gamma_{FluidRMHD}/\omega^*$ | $\nu_e/\omega^*$ | $\gamma_{0Fluid}/\omega^*$ | $\hat{\beta}$ |
|---|---|---|---|---|---|---|
| JET 79697 | 8 | 0.18 | 0.00011 | 0.4 | 0.048 | 730 |
| JET 53062 | 3 | 0.07 | 0.00003 | 0.07 | 0.030 | 710 |
| JET 82585 | 3 | 0.17 | 0.00005 | 0.3 | 0.037 | 670 |
| DIII-D 153764 | 14 | 0.13 | 0.00034 | 1.2 | 0.070 | 700 |
| ASDEX 30710 | 12 | 0.18 | 0.00026 | 0.4 | 0.064 | 730 |
| ASDEX 30721 | 12 | 0.12 | 0.00009 | 0.15 | 0.046 | 730 |

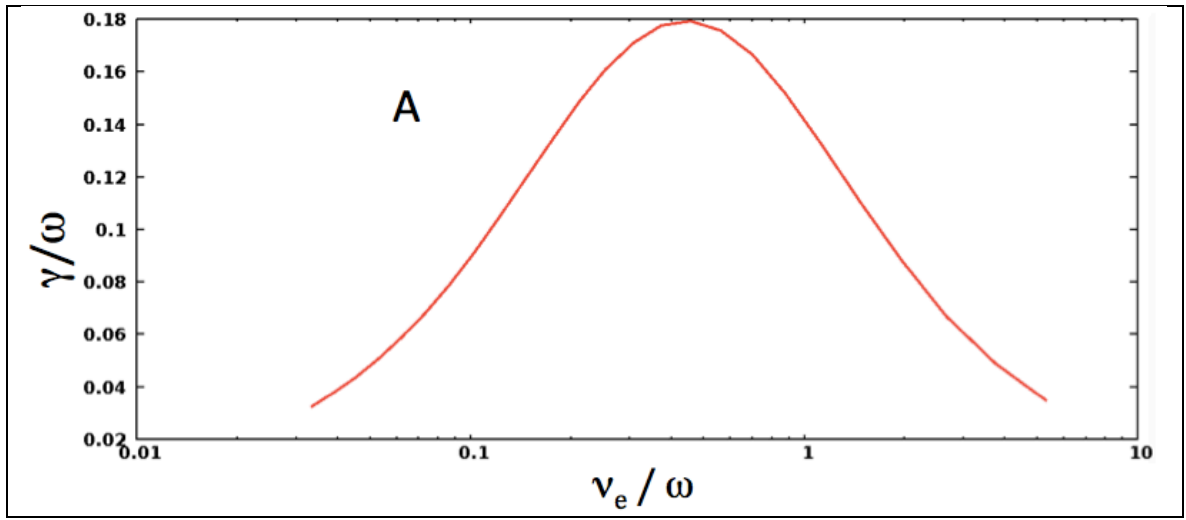

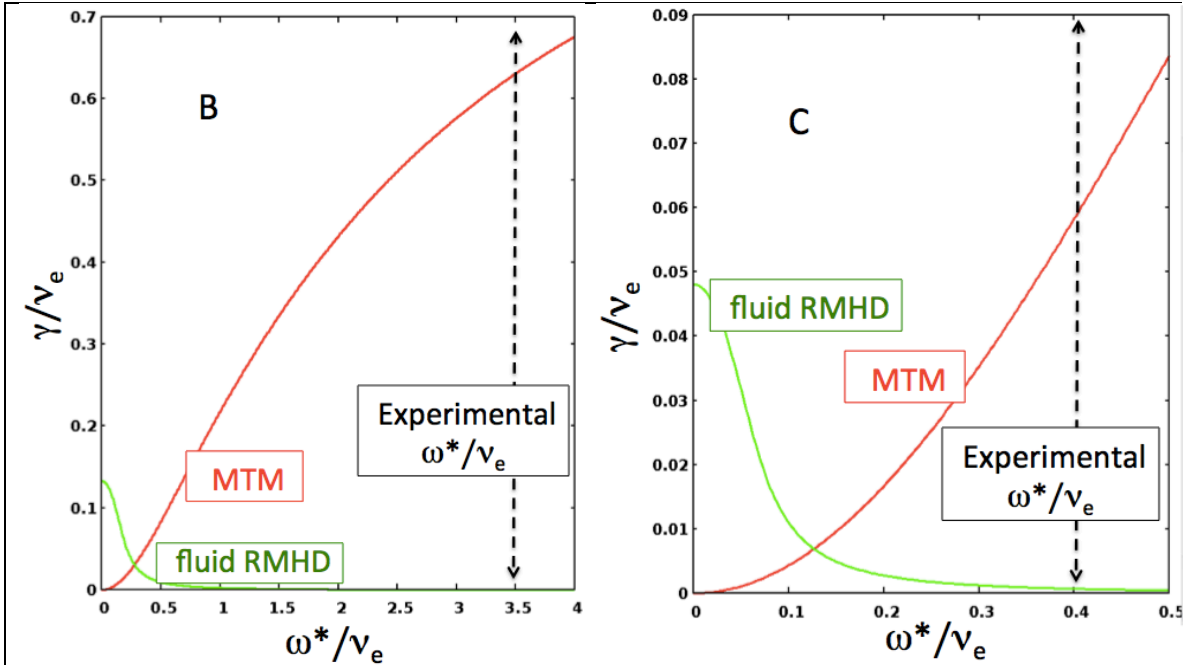

Figure 3. In A, the analytic MTM dispersion relation is plotted, showing the range of collision frequency $\nu_e/\omega$. The instability is driven over a wide range of collisionality. It is significantly driven, with $\gamma/\omega$ of ~ 1/3 of the maximum, over a very large range, from about $0.06 < \nu_e/\omega < 3$. In B and C, we show growth rate trends for the experimental $\nu_e$, as $\omega^*$ is increased. The JET shot in B has moderate collisionality (79697 in B), and a much more collisional case is shown in C (82585). The fluid resistive MHD modes become strongly *stabilized* for the large $\omega^*$ typical of pedestals, whereas the MTM become strongly *destabilized*.

In addition to the huge size of $\omega^*$, there is another important parameter that favors MTM over fluid RMHD modes in a pedestal. This is the ratio of the magnetic skin depth to the current channel width around the rational surface where reconnecting current can arise. Recall that, physically, the skin depth is the distance over which magnetic shielding can occur. In the strong shielding regime, the driving region of fluid RMHD modes is screened from the crucial reconnecting region, as first described in ref [53]. The dimensionless parameter delineating this regime is when $\hat{\beta} = \beta_e(L_s/L)^2 \gg 1$. The $\hat{\beta}$ is much larger in pedestals than in the core because of the very steep gradients (very small L), and is often $> 10^2$ (see Table 2). This *additional* strongly stabilizing effect for fluid RMHD modes is not taken into account in the results in Fig. 3 and Table 2, since simple analytic formula are not available for this effect. On the other hand, for MTM, the effect is the opposite. The region of maximum kinetic instability drive, which arises inside the current layer, shields itself from the regions of kinetic damping, leading to an *even stronger* instability. As indicated in ref [53], when $\hat{\beta}$ is large, the MTM is expected to always be unstable.

The discharges in Table 2 span a considerable range of parameters for ELMy H-modes, and show that MTM predominate by a wide margin even without including the stabilization of fluid RMHD modes by large $\hat{\beta}$. The large magnitude of $\hat{\beta}$ further strengthens this already strong conclusion.

The analytic arguments in the literature also clearly show that MTM are expected to often be unstable for pedestal parameters, given that they fall in the appropriate range of $\nu_e/\omega^*$, and, that $\hat{\beta}$ is large.

Analytical arguments further show that MTM in the large $\hat{\beta}$ regime are primarily magnetic, with a relatively small electrostatic potential. Hence, for MTM, the main source of transport is from electron motion along perturbed field lines, rather than perturbed ExB motion. This is exactly the opposite of ideal MHD-like modes such as the KBM. So MTM have predominantly electron thermal transport. There is little electron particle transport because of the essential requirement that the transport be ambipolar, and ion parallel motion is much slower. We consider this question further in the kinetic analysis of section VI.

Crucially, the MTM has a very different fingerprint from the MHD-like modes, which greatly assists in distinguishing the modes based on their transport. One can examine the apparent effects of the growing fluctuations on various pedestal profiles, such as $T_e$, $n_e$, $T_i$ and $n_z$. We examine *all* these channels for the DIII-D shot 153674 considered in section V, confirming that the mode is indeed an MTM.

A commonplace observation is that pedestals reach a quasi-steady state for a significant time before an ELM. The magnetic fluctuations that appear to be MTM usually grow to maximum amplitude when the pedestal $T_e$ saturates. In this phase, both the gradients are largest and fluctuation amplitude are strongest, so we expect the largest transport losses occur in this time period.

*In summary, observed magnetic signals on multiple devices, which apparently have little effect on the density profile, are direct evidence for the presence of MTM in pedestals.*

We discuss two DIII-D cases in section VI, where thorough analysis enables identification of the fluctuations as MTM especially unequivocally, using extensive gyrokinetic simulations together with the fingerprint concept.

We close by noting that MTM are relatively resistant to ExB shear suppression, and hence, they are a candidate to cause pedestal transport. This is because the instability drive occurs in a radially narrow layer near a rational surface. This region is far narrower than the typical radial scale of electrostatic modes such as ITG modes. Hence, MTMs feel the effect of a radial shearing in velocity much less than ITG modes. This basic concept is reflected in analytic shear suppression theories: the criterion for suppression is $(\Delta x/\Delta y)\, \gamma_{ExB} > \gamma$ where $\Delta x$ and $\Delta y$ are the decorrelation lengths in the radial and poloidal directions [58-60]. For electrostatic modes in the core, $\Delta x/\Delta y$ is, reasonably, taken as $\sim 1$. But for MTM in pedestals $\Delta x/\Delta y$ is very small. The radial scale can be estimated from the analytic literature, and for large $\hat{\beta}$, is roughly one and a half orders of magnitude less than $\Delta y \sim 1/k_y$. Together with the growth rate estimates from analytic theory (as in Fig 3A) $\gamma \gg (\Delta x/\Delta y)\, \gamma_{ExB}$, so these modes should not be suppressed. Local gyrokinetic simulations for the DIII-D cases here corroborate that $\Delta x/\Delta y$ is very small (using the linear eigenfunctions, see section V and VII), and that $\gamma \gg (\Delta x/\Delta y)\, \gamma_{ExB}$. Global simulations with GENE also confirm the conclusions from these local expressions (see section VII).

Hence MTM should not be suppressed by velocity shear, and, should be considered as a candidate to cause pedestal transport.

Electron Temperature Gradient (ETG) modes

These instabilities are characterized by perpendicular scales shorter than an ion gyroradius, $k_\perp \rho_i > 1$ [10,11,14-19,61]. In the limit of large $k_\perp \rho_i$, ions are adiabatic [58]. In fact, the adiabatic response of the ions implies that all ion transport fluxes vanish. Hence ETG produce transport almost solely in the electron channel, and as above, due to quasi-neutrality, there is only electron thermal transport. The very large magnitude of the growth rate of ETG modes easily exceeds the shearing rate. Previous simulations in pedestals find that they can cause significant transport, so they should be considered candidates for pedestal transport.

Ion Scale Electrostatic modes

These modes are usually dominant in the core. They have $k_y \rho_i < 1$. In a pedestal of todays experiments, these modes are usually suppressed by ExB shear. Nonlinear results, so far, indicate that for large enough $\rho^*$ (as on ASDEX-U, DIII-D, JET-C and JET-ILW at low B [19,20,42]), suppression is strong enough that these modes do not play a dominant role. The modes are expected to be important on JET-ILW at high field and/or current [19,20,62], and potentially, some other cases with unusual profiles of $E_r$ or plasma parameters.

The main results of this paper persist even if ITG/TEM, with their characteristic fingerprints, may be active. The fingerprints of ITG/TEM modes are relatively complicated, compared to those from MHD-like modes, MTM and ETG. We have recently completed an extensive analysis of electrostatic modes in pedestals, using both analytic and numerical methods. Except what is pertinent to the fingerprint concept, these results on stability, mode structure, etc., will be presented elsewhere [63]. The general characteristics are manifested in gyrokinetic simulations for the two DIII-D pedestals considered in section VIII (unsurprisingly), so we may regard those two examples as paradigmatic.

The TEM modes dominate when the ion temperature gradients are relatively weak, so that ITG modes are weak or absent. In gyrokinetic runs for an ASDEX-U pedestal, a form of ETG emerges as the dominant electron mode. Even when $k_y \rho_i < 1$, these modes have high radial wave numbers so that $k_\perp \rho_i > 1$. For the DIII-D runs, since the mode frequency $\omega \leq \omega_b$ (the electron bounce frequency), qualifying them as a branch of TEM. We refer to them as TEM/ETG modes. Despite their high growth rates, a mixing length estimate of diffusivity $\gamma / k_\perp^2$ is *extremely* small (because $k_\perp$ is large): about three orders of magnitude lower than the heat diffusivity estimated empirically from power balance. We do not pursue these modes further.

We now turn to ITG modes that will grow when there are substantial gradients of $T_i$. Being driven mainly by $T_i$ gradients, unsurprisingly, these modes have $\chi_i > \chi_e$, or sometimes, $\chi_i \sim \chi_e$. The same ExB fluctuations that cause ion temperature transport also cause impurity transport, with $D_Z/\chi_i \sim 1$.

A pinch is sometimes found with pedestal ITG/TEM modes (negative $D_e$ in our definition). However, we always find that such negative $D_e$ are small in magnitude, $|D_e/(\chi_i + \chi_e)| < 10\%$. For energetic reasons, we expect that inward pinches must be weak in comparison to outward energy transport. In order to grow, a mode *must* reduce the free energy of the equilibrium by the same amount as the mode energy increases. An inward pinch *increases* the free energy in the density gradients; the free energy in the temperature gradients must be *reduced* even more to overcome this. The temperature flattening (decreasing the free energy), in addition to providing free energy so that a pinch enables density gradients, also supplies the free energy for; 1) the increased kinetic energy from ExB motions, and 2) collisional dissipation in fluctuations. Hence, when there is a pinch from an ITG/TEM mode, the reduction in free energy due to temperature gradient reduction must be quite a bit larger: in other words, $\chi$ must be quite a bit larger than $|D_e|$. No wonder that the simulations find $|D_e/(\chi_i + \chi_e)| < 10\%$ when $D_e < 0$.

Alfven Eigenmodes (AE/MTM)

As noted above, a unique characteristic of the pedestal is the huge size of $\omega_e^*$, so that it can be of order or greater than Alfven frequencies. A corollary of this is that

Alfven Eigenmodes (Toroidal Alfven Eigenmodes and Kinetic Alfven Eigenmodes [64-66]) can have the same frequency as other modes, in particular MTM, so that hybrid AE/MTM modes can arise. We find such modes in local linear calculations with GENE for the DIII-D cases here, but they have low growth rates, and have the same fingerprint as MTM. Global simulations find that these modes become primarily magnetic and MTM-like, so we consider them to be a variant of MTM.

**Section V: Application to two DIII-D cases**

We will now analyze two distinct DIII-D shots by a combination of analytic theory (sections IV and VI) and simulations; simulation results fully corroborate analytic predictions.

The inter-ELM behavior of profiles and Quasi Coherent Fluctuations (QCF) [6] were extensive diagnosed for the identical pair of shots 153674/153675 (hereafter referred to as 153674/5). We will identify the QCF as arising from an MTM, and electron energy is dominated by MTM and/or ETG.

We also consider shot 98889, where an extremely valuable transport analysis was published [25]. In addition, a magnetic spectrogram reveals a high frequency QCF similar to 153674/5. We identify the energy transport agents as MTM, which is also the source of the QCF.

<u>DIII-D Shot 153674/5- experimental observations and their implications</u>

A magnetic spectrogram of the QCF is shown in figure 4A. In some ELM periods, there is a second QCF as well, however, it was generally subdominant to the higher frequency (f) mode, and was not as reliably correlated with $T_e$ evolution.

The dominant QCF has f ~140kHz, in the electron direction (the other QCF had f ~100kHz). It is found on magnetic diagnostics and on BES; the latter is used to identify $k_y$ to be ~ 0.18 -0.2 cm$^{-1}$. To identify the mode by its frequency f, it is crucial to establish f in the plasma frame. In figure 4B, we plot the Doppler shift from the measured $E_r$ and $k_y$ and $\omega_e^*$. Even the *maximum* Doppler shift ($\omega_{ExB}(r)$) is much less than the lab frequency with the implication that *f must be in the electron direction in the plasma frame, wherever the mode is "located".* The measured f is roughly consistent with $\omega_e^*$ for the experimentally inferred position of the mode, but with a large uncertainty.

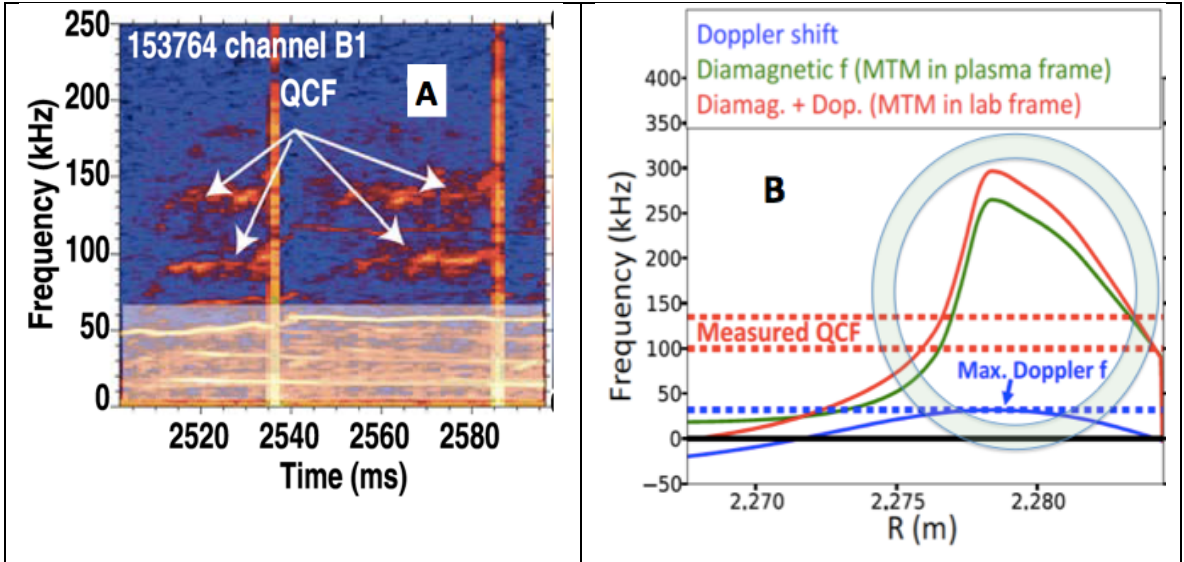

Figure 4. DIII-D shot 153674 a) Magnetic spectrogram of the experimentally observed fluctuations, showing measured QCF and nonlinear simulation frequencies b) From experimental profiles, frequencies f for: Doppler shift ($\omega_{ExB}$), $\omega_e^*$, and the QCF. The circle shows the experimentally inferred maximum amplitude position of the fluctuation.

As emphasized in the published analysis of this shot [6], the growth of the magnetic signals is strongly correlated with the evolution of the $T_e$ gradient (see figure 5A). This strongly suggests that the QCF is driven by it, and/or, causes substantial transport in that channel. It is also seen that the QCF saturates at about the same time as the $T_e$ gradient, greatly strengthening the physical link between them.

For mode identification, it is just as significant that the growing QCF has no apparent effect on the evolution of the density gradient; this gradient saturates well before the onset of the QCF (Figure 5B). When seen in the backdrop of the fingerprint concept, the observations, though consistent with an MTM, rule out an MHD-like mode that would be expected to affect the density gradients as well.

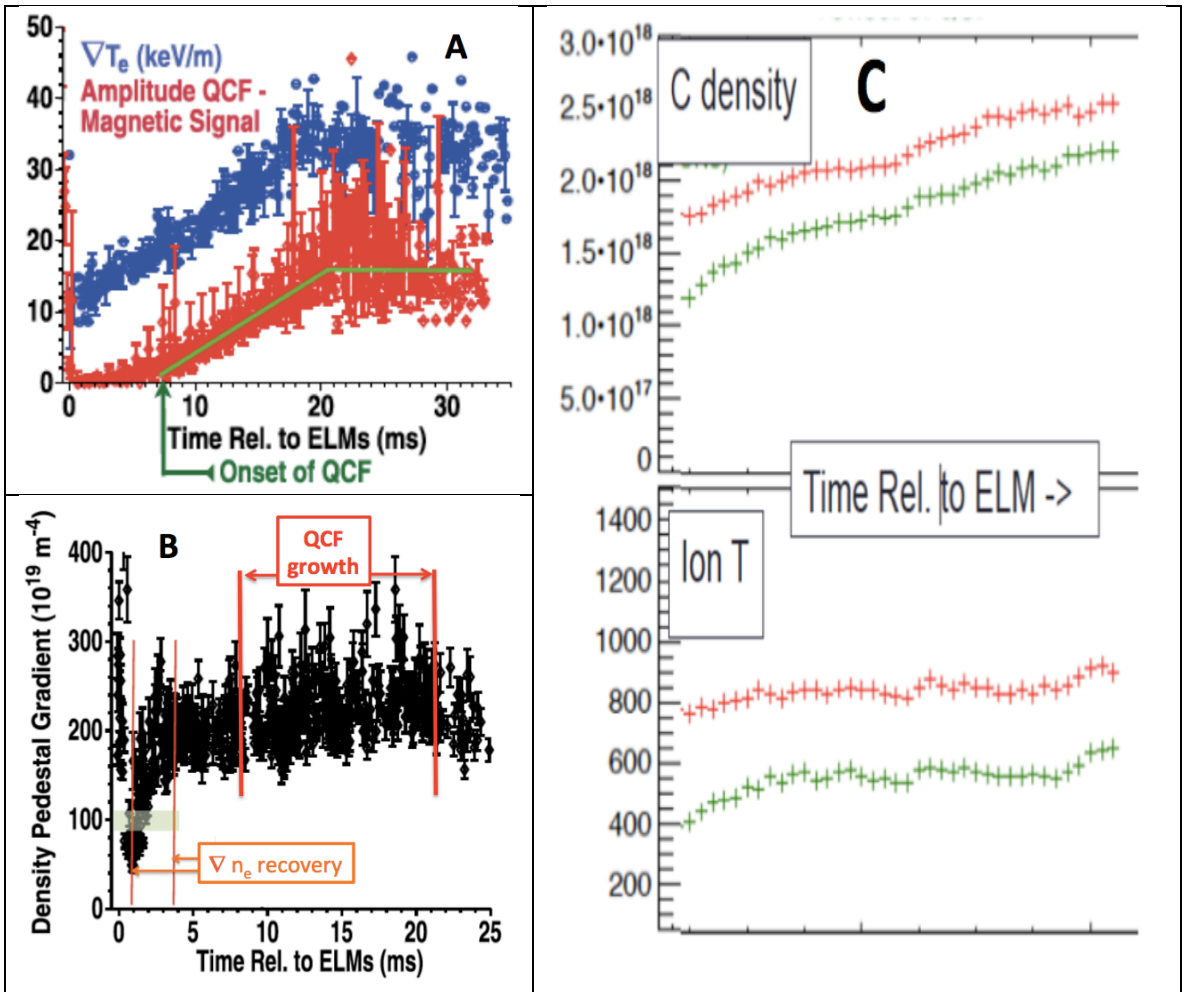

Figure 5. DIII-D shot 153674. In A), the inter-ELM evolution of $T_e$ gradient and QCF amplitude showing strong correlation, expected for an MTM. In B), the evolution of the electron density gradient, which saturates much more quickly, and is unaffected by the growing QCF  In C) the evolution of $T_i$ and $n_C$ after an ELM, for a typical ELM cycle, showing no discernable affect of growing QCFs- consistent with MTMs, but not KBMs. Note the difference between values on two cords is proportional to the average gradient between them. Hence, the gradient of $T_i$ and $n_C$ between the chords is apparently unaffected by the growing QCF.

The fingerprint concept led us to examine the inter-ELM evolution of other channels: $T_i$, and Carbon density $n_C$. Pedestal measurements of $T_i$ and the impurity density $n_C$ are shown in figure 5C, for a representative inter-ELM period. Data from two separate chords in the pedestal are plotted. The spatial separation between the chords does not change in time, so the difference in the signal between the chords is proportional to the average plasma gradient of the quantity between them. This difference is very nearly constant in time for both $T_i$ and $n_C$. One would expect that an MTM with growing amplitude would not produce a change in the transport in the

$n_e$, $T_i$, and $n_C$ channels, so this pattern is consistent with the fluctuation being an MTM.

In the phase when the amplitude of the perturbation is increasing, the transport it is producing should be increasing as well. If the perturbation were a KBM, which has extremely stiff transport, the KBM evolution would rapidly produce transport that is so large that it prevents a further increase in the gradients. As it grows, it would be producing increasingly large, and comparable, transport in all channels. Necessarily, the transport would eventually be strong enough to affect the pedestal gradients. However, during the phase when the magnetic fluctuation amplitude is strongly increasing, the observed gradients of $n_e$, $T_i$, and $n_C$ show no evidence at all of strongly increasing transport; quite the contrary, those channels are behaving as if the transport which controls them is totally unaffected by the growing mode that is associated with the magnetic fluctuation. The electron temperature profile, eventually, is apparently affected, in that the $T_e$ gradient stops increasing- exactly as one would expect from increased transport in the $T_e$ channel. For the case of a weakly driven electron density channel, one would expect that a growing MHD-like mode would affect the $n_e$ profile even more strongly than it affects the $T_e$ profile (perhaps, even, by causing the $n_e$ gradient to decrease), but no affect of the growing mode associated with the magnetic fluctuation is apparent. The experimentally observed pattern is inconsistent with the possibility that the observed magnetic fluctuation is due to an MHD-like mode that evolves to marginal stability by significant transport in all channels. On the contrary, the pattern is completely consistent with the evolution of an MTM which eventually saturates the $T_e$ profile evolution, and does not affect other channels.

We note that the pattern above is repeated for a very large number of ELM cycles in this shot; it is quite typical.

Similarly, on the basis of the fingerprint of ITG/TEM (expected diffusion in $T_i$ and $n_{Carbon}$ channels, and, frequency in the plasma frame), they can also be excluded (though such primarily electrostatic modes would not be expected to give a strong magnetic signature).

Hence, we conclude that the observed QCF is an MTM. This is the first strong identification of MTM in pedestals based on both frequency, and, behavior in multiple transport channels.

We now turn to GENE simulation results to corroborate the analytic considerations of section IV and VI.

*Local simulations* with GENE do indeed find that MTM are unstable at radial locations in the middle pedestal, for a wide range of $k_y$, including the experimental value (See Fig. 6 for mode properties indicated in the following). The modes are primarily magnetic and are good candidates to produce transport (Sec. IV) since their growth is strong enough to avoid shear suppression: (the growth rate $\gamma$ greatly exceeds the effective ExB shearing rate $\gamma_{ExB}$ ($\Delta x/\Delta y$) (where ($\Delta x$ and $\Delta y$ are estimated from the linear eigenfunction as $k_y/k_\perp$. (Note that the profile of $E_r$ obtained from the pedestal CXRS system is used to estimate $\gamma_{ExB}$. Since $\gamma_{ExB}$ is a derivative of the measurements, it may have significant error bars, but we use it as the best available estimate.)

The MTM mode frequency is close to $\omega_e^*$ (computed from the gradient of $n_eT_e$). They are strongest in mid-pedestal, where $\omega_e^*$ is largest; at the observed $k_y$ value, however, the lab-frame frequency found by GENE is typically higher than the QCF value by ~ 2. As we will see, nonlinear effects considerably reduce the difference.

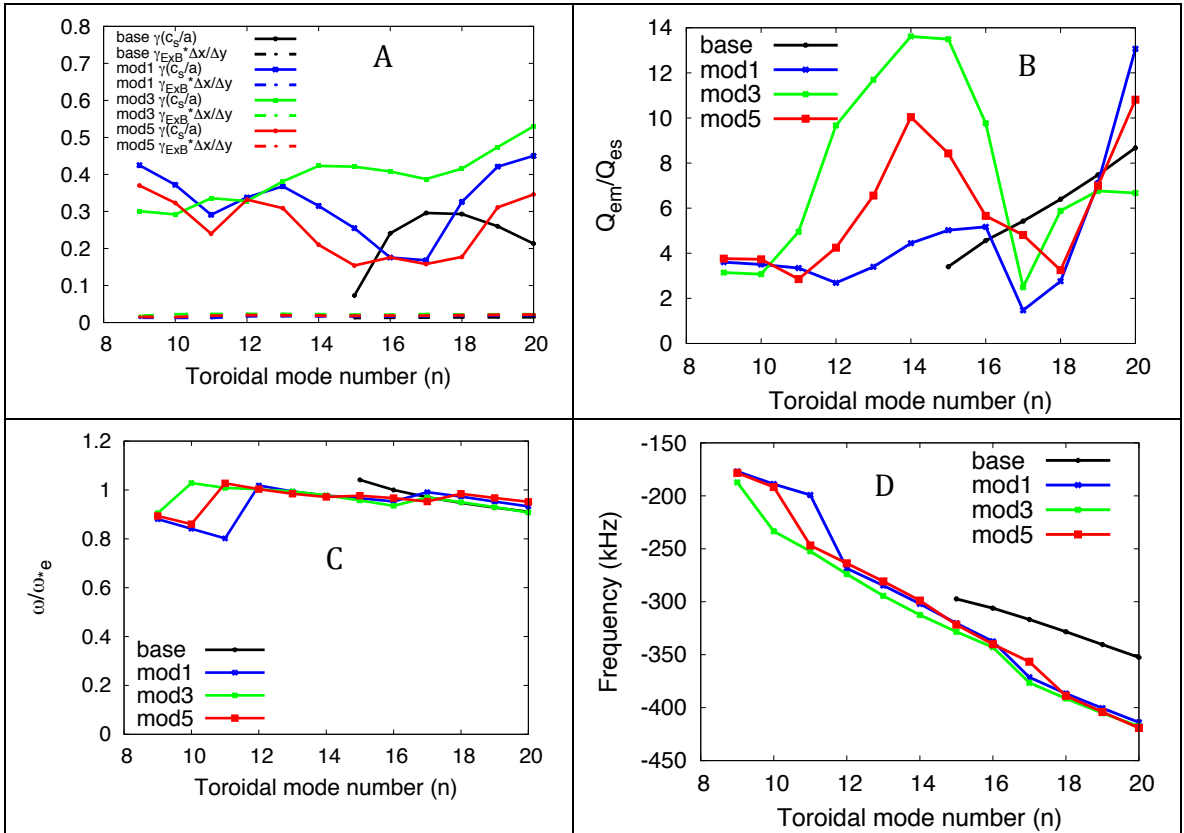

Figure 6. Properties of representative MTM found by GENE for 153674/5. (A), growth rates $\gamma$ are much larger than the effective shearing rate $\gamma_{ExB}$ ($\Delta x/\Delta y$) , so the modes should avoid suppression. In (B) the quasi-linear heat flux is mainly from magnetic fluctuations, so they are MTM. In (C) the frequency in the plasma frame is close to $\omega_e^*$, computed from the gradient of $n_eT_e$. In (D), frequencies in the lab frame, including the addition of the measured $E_r$ Doppler shift, are shown.

Local linear runs also find KBM close to marginal stability near the top of the pedestal, in agreement with previously published analysis [6] with the gyrokinetic code GS2 [67]. Based upon the GS2 instability, and the good general agreement of the pedestal profiles and evolution with the EPED model, ref [6] suggested that the

QCF might possibly be a KBM. However, as we have described in section II, marginal stability of profiles to MHD-like modes does not imply that such modes are the dominant energy loss processes.

The KBM is particularly prominent in the base case, so we display mode properties for this profile. As in ref [6], we examine whether KBM are close to marginal stability by artificially increasing β in the simulation, and doing a $k_y\rho_s$ scan. As can be seen in figure 7, increasing β from 1.1 to 1.2 times the nominal value ($\beta_{nom}$) results in large and rapid increases in the growth rate in the range $k_y\rho_s < 0.3$. This is strongly indicative that this is a KBM instability, which, at least in a local analysis, is fairly close to marginal stability for $\beta_{nom}$. The growth rate exceeds the shearing rate for β ~ 1.2 -1.3 $\beta_{nom}$, so it could cause strong transport. To verify that this is an MHD-like mode, we compute the eigenfunction averaged parallel electric field $<E_{||}>$. The cancellation of inductive and electrostatic components of $E_{||}$ becomes quite strong for β > 1.1 $\beta_{nom}$. The frequency of this mode is slightly in the electron diamagnetic direction, but is only a small fraction of $\omega_e^*$, (typically $\omega \sim \omega_e^*/4$) for the $k_y$ observed on the outboard mid-plane. Most importantly, in the lab frame, and including the measured Doppler shift, f ~ 20-35 KHz, or about 4 times lower than the dominant QCF value.

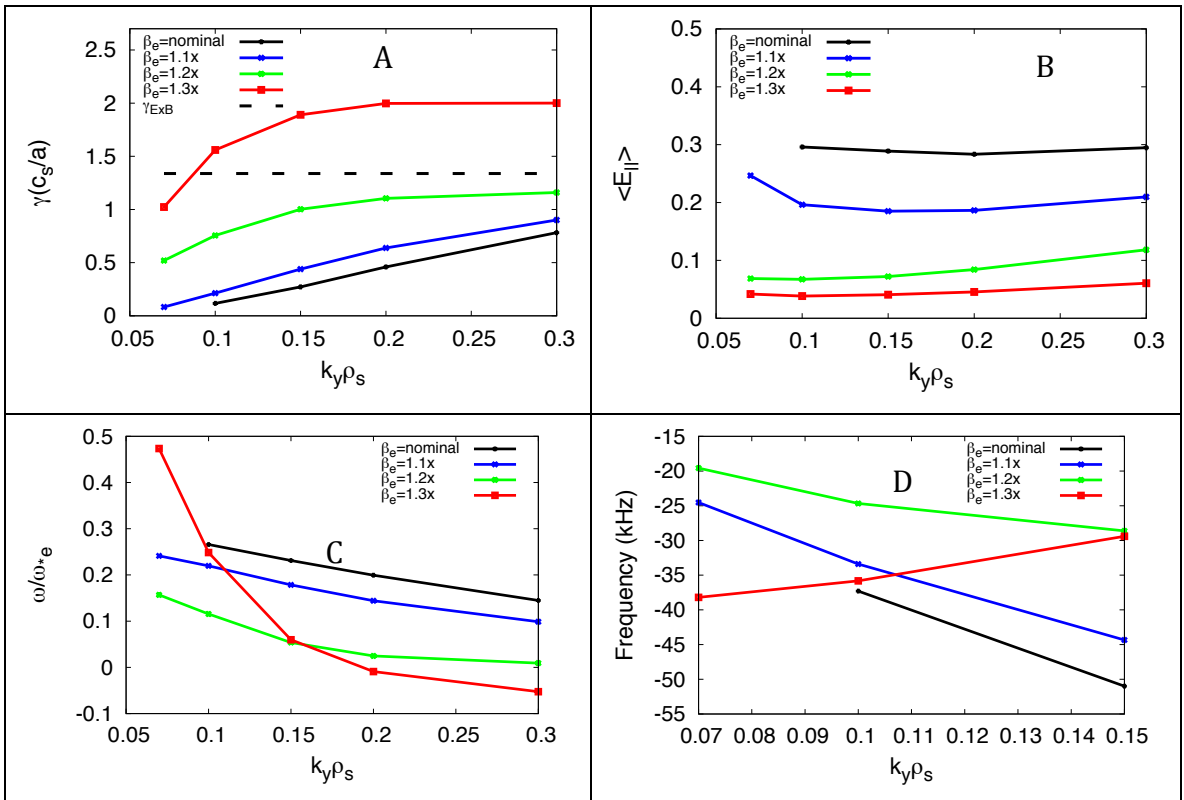

Figure 7. In (A), growth rates γ increase strongly with β when it is increased by a factor > 1.1. Also, γ becomes much larger than $\gamma_{ExB}$ so the modes should avoid

suppression. In (B), the average $E_{||}$ also becomes small, the signature of MHD-like modes. In (C) the frequency is considerably smaller than $\omega_e^*$, though it is in the electron direction. In (D), frequencies in the lab frame, including the measured $E_r$ Doppler shift, are shown, and are 3-4 times less than the measured QCF. (Note the definition of $k_y\rho_s$ used by GENE is not the value at outboard midplane, unlike GS2 [67], but is ~ 2-3 times higher for these cases.)

The local linear results indicate that both MTM and KBM can overcome velocity shear suppression. As discussed in section VII, the MTM are much more robust in a different but important sense: they were unstable for a large enough range of $\Delta k_x$ so that unstable modes could likely "fit" into a box of width w of the pedestal: $\Delta k_x w > 2$. It would seem, then, that only MTM might actually arise in the finite width pedestal. Global runs, described below, bear this out. (However, we must add the caveat that the version of GENE that was used did not include some MHD effects that are absent in lowest order gyrokinetics, such as the kink term and vacuum magnetic boundary conditions; inclusion of these might lead to a global MHD instability.)

Electron scale ETG modes were also unstable for these profiles. As indicated in section VII, the quasi-linear transport fingerprint, as expected, had almost exclusively electron thermal transport.

Other instabilities were found in local linear runs: TEM/ETG modes and MTM/AE modes. However, as discussed in section VII, these are apparently too weak to avoid shear suppression, cause significant transport, or produce significant signals.

The transport fingerprints, calculated from GENE simulations, are displayed below. The local quasi-linear results for the MTM, are shown in Fig.8. The ratios of the diffusivities for $T_i$, $n_e$ and $n_{Carbon}$ to $\chi_e$ are found to be small (~ 0.1).

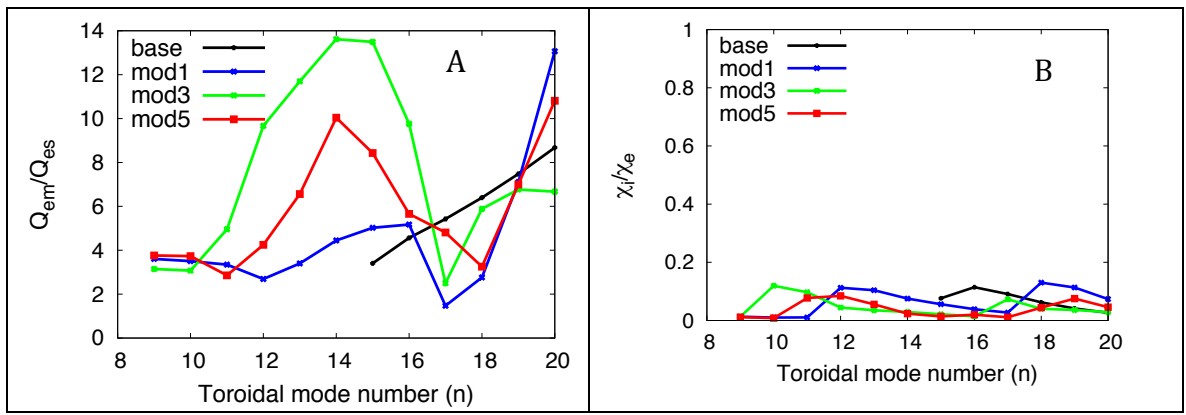

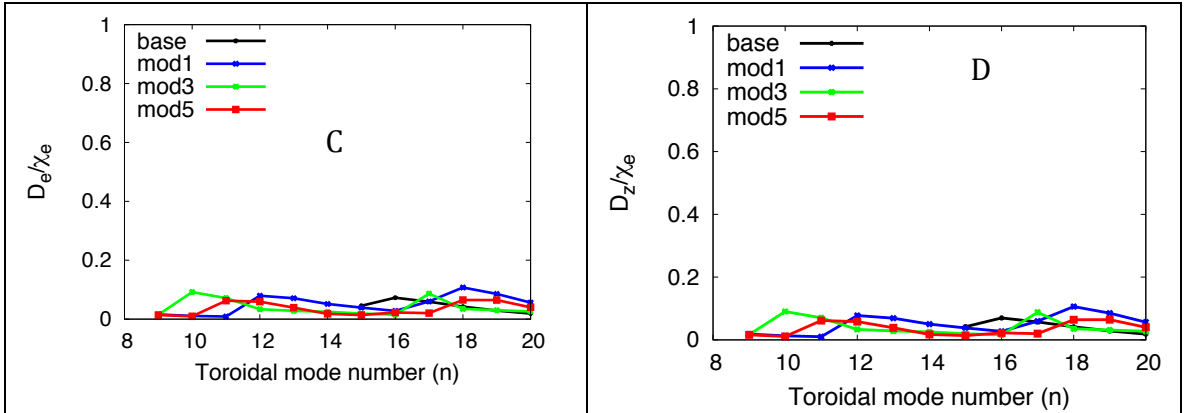

Figure 8. Transport characteristics for local linear MTM for DIII-D shot 153674. In (A), we see that the flux from magnetic fluctuation dominates, as expected. Thus, other diffusivities are small compared to $\chi_e$: for $T_i$ (B), $n_e$ (C) and $n_{Carbon}$ (D).

These fingerprints (calculated from local linear simulations) are as expected, and similar to ones obtained from global linear, and nonlinear, results (see section VII).

The fingerprints for the local linear KBM found above are in good agreement with the analytical expectations, as shown in fig 9.

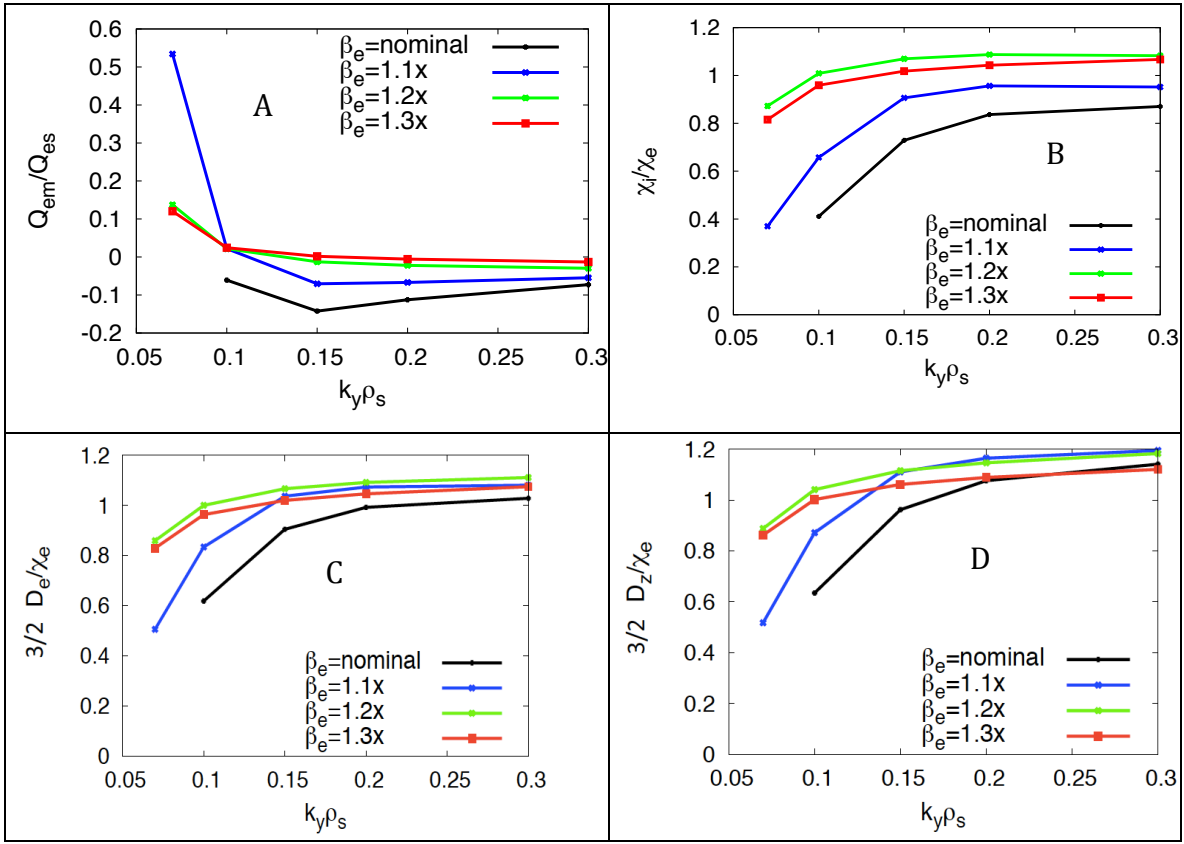

Figure 9. Transport characteristics for local linear KBM for DIII-D shot 153674. In (A), we see that the flux from electrostatic fluctuation dominates, as expected for an MHD-like mode. Thus, all diffusivities are comparable, with ratios close to the analytic expectations: for $T_i$ (B), $n_e$ (C) and $n_{Carbon}$ (D).

We now turn to *linear global simulations* with GENE; these include the full variation of the pedestal profiles, including the $E_r$ profile from CXS. *The only instabilities that survive are the MTM (Fig.10).*

The global instability spectrum (over the toroidal mode number n) has quite a surprising structure and sensitivity. Only isolated toroidal modes are excited; a few unstable n numbers, are, usually, separated by several stable n. *Such instability spectrum should lead to discrete QCFs, as observed,* rather than the broad-band turbulence expected from an ordinary spectrum.

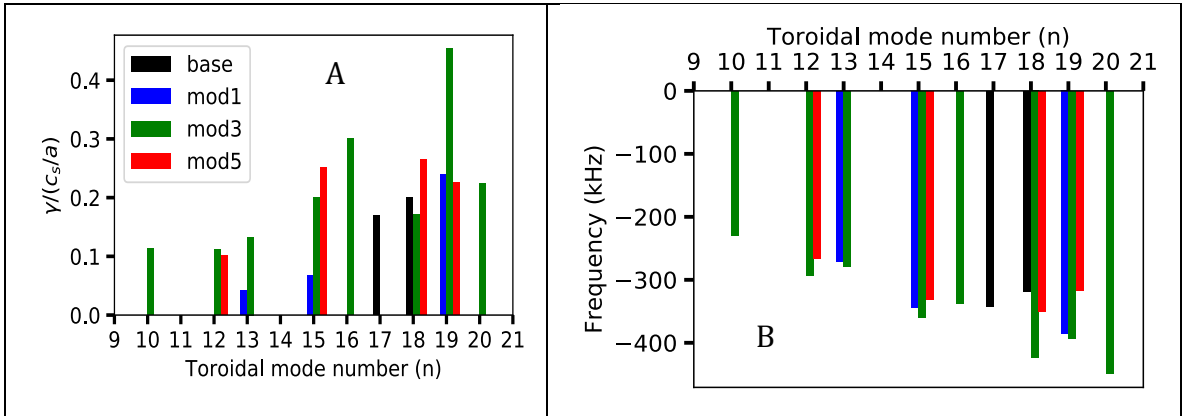

Figure 10. Global linear spectrum of MTM instabilities for 153674/5. Instability occurs at discrete n numbers, usually separated by stable modes. This would lead to Quasi-Coherent modes, as observed experimentally. The spectrum is very sensitive to the profile. Modes with the observed n number (~ 13) have a frequency about 2x the QCF.

The unstable n values differed significantly for the profiles. One profile nearly matched the observations: mod 5. It had a clearly dominant instability for $k_y$ within ~ 10% of the measured value for the QCF.

In addition to the dominant instability, many profiles (including mod 5) sustain a second weaker MTM with lower n and frequency; again qualitatively matching the observed magnetic spectrogram.

The global MTM usually have maximum amplitude in the region of steep profile gradients, where $\omega_e^*$ is large; consequently, their linear mode frequency in the lab frame is larger than the QCF frequency by a factor ~ 2.

Nonlinear simulations find a strong downshift in the frequency, to ~ 200 KHz, which is much closer to the values observed, but still too high by ~ 40%. Results from several nonlinear simulations are shown in the table 3.

TABLE 3 Nonlinear MTM simulation results for profiles of DIII-D shot 153764/5

| Case | Toroidal number n | Heat flux at mid pedestal $\rho = 0.98$ (MW) | Linear frequency (kHz) | Nonlinear frequency (kHz) | Nonlinear frequency downshift |
|---|---|---|---|---|---|
| base | 17 | 0.5 | -343 | -285 | 17% |
| mod1 | 15 | 2.3 | -341 | -238 | 30% |
| mod3 | 13 | 0.28 | -279 | -171 | 39% |
| mod5 | 15 | 0.8 | -331 | -197 | 40% |

Nonlinear ETG simulations at mid pedestal found a very strong variation in heat diffusivity with the profile( Table 4):

TABLE 4: Nonlinear ETG heat fluxes at r = 0.982

| Case | $Q_e$ (MW) |
|---|---|
| base | 0.89 |
| mod1 | 4.97 |
| mod3 | 3.35 |
| mod5 | 1.39 |

Nonlinear MTM results for mod 5 (with a $k_y$ spectrum closest to observations) find that the MTM account for roughly 0.8 MW of heat loss. Together with the ETG transport of 1.4 MW, this adds to ≈ 2.2 MW. This is in reasonable agreement with power balance (total losses are ≈ 3 MW) especially considering that there must be additional losses due to ion neoclassical transport and ELMs. Details of the MTM nonlinear runs are given in section VII. Given all the profile sensitivities and model uncertainties noted in section III, we conclude that, within the error bars, a combination of MTM plus ETG is capable of accounting for anomalous inter-ELM energy losses in this shot.

Overall, simulation results for the frequency and spectrum of the instabilities match reasonably well with MTM, and do not match with KBM.

<u>DIII-D shot 98889 -experimental observations and their implications</u>

For the shot DIII-D 98889, an extremely valuable previous transport analysis [25] revealed:

    a) The inferred $\chi_e$ is about twice $\chi_i$

    b) The latter $\chi_i$ is roughly consistent with neoclassical or paleoclassical expectations.

    c) The inferred electron diffusivity *$D_e$ is an order of magnitude lower than* $\chi_e$

By arguments delineated in section II, the smallness of $D_e$ (section II.4), and also the relative smallness of the ion anomaly $\chi_i - \chi_{neo} \ll \chi_e$ (section II.2), preclude MHD-like modes from being the dominant energy loss mechanism for 98889.

We expect that ITG/TEM modes are probably strongly suppressed in this pedestal; in any case, the relative smallness of the inferred ion anomaly $\chi_i - \chi_{neo} \ll \chi_e$, is inconsistent with them dominating the energy losses.

The only consistent candidates left to account for anomalous energy transport are, therefore, MTM and/or ETG.

Note that a magnetic spectrogram for this shot (see Figure 11), shows QCFs that are correlated with the ELMs, and which grow in strength as the inter-ELM phase proceeds. The upper bands have f ~ 180 and 220 kHz in the electron direction.

Though $k_y$ could not be determined experimentally, it can be bounded. Magnetic perturbations exponentially decay in the vacuum, with an e-folding length scale of ~ 1/ $k_y$. The magnetic diagnostic was located ~ 20 cm away from the plasma edge, so one can estimate that the largest $k_y$ that might be detected corresponds to $k_y$ ~ 0.5 cm$^{-1}$ (or poloidal m ~ 80). This bound allows us to compute an upper bound on the Doppler shift from the measured $E_r$, which is only about 40% of the observed f. Hence, the signal must be strongly in the electron diamagnetic direction in the plasma frame, which is consistent with MTM, and inconsistent with KBM.

It is legitimate, then, to conclude that these fluctuations are likely MTM.

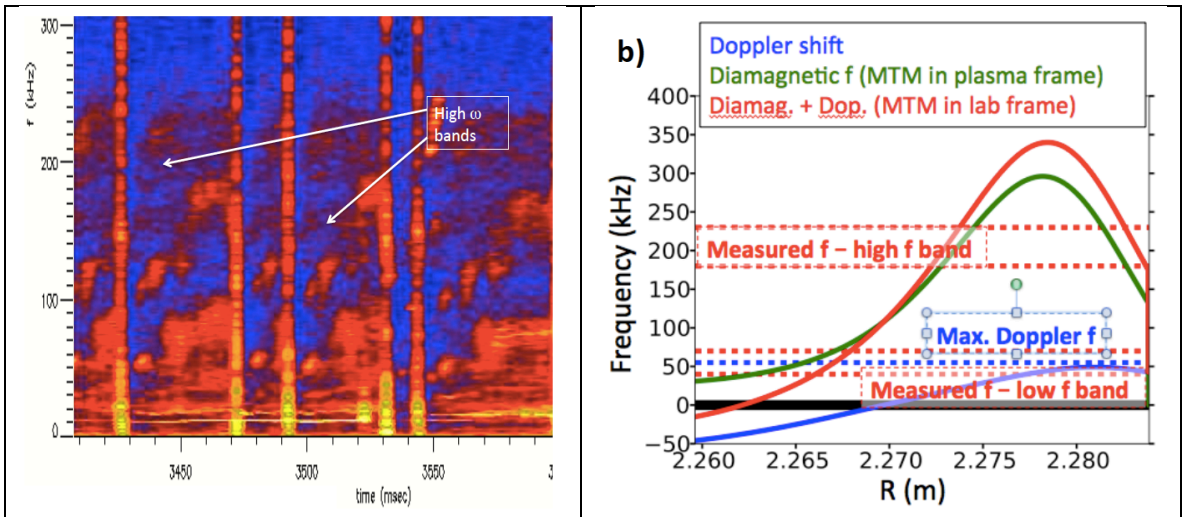

Figure 11. DIII-D shot 98889 A) Magnetic spectrogram showing observed QCF B) From experimental profiles, frequencies f for: Doppler shift ($\omega_{ExB}$), $\omega_e^*$, and the QCF.

GENE simulations for the shot 98889 corroborate the preceding conclusions.

In local linear simulations, the MTM are found at most radii in the mid to upper pedestal. A typical example, with its relevant properties, is shown below.

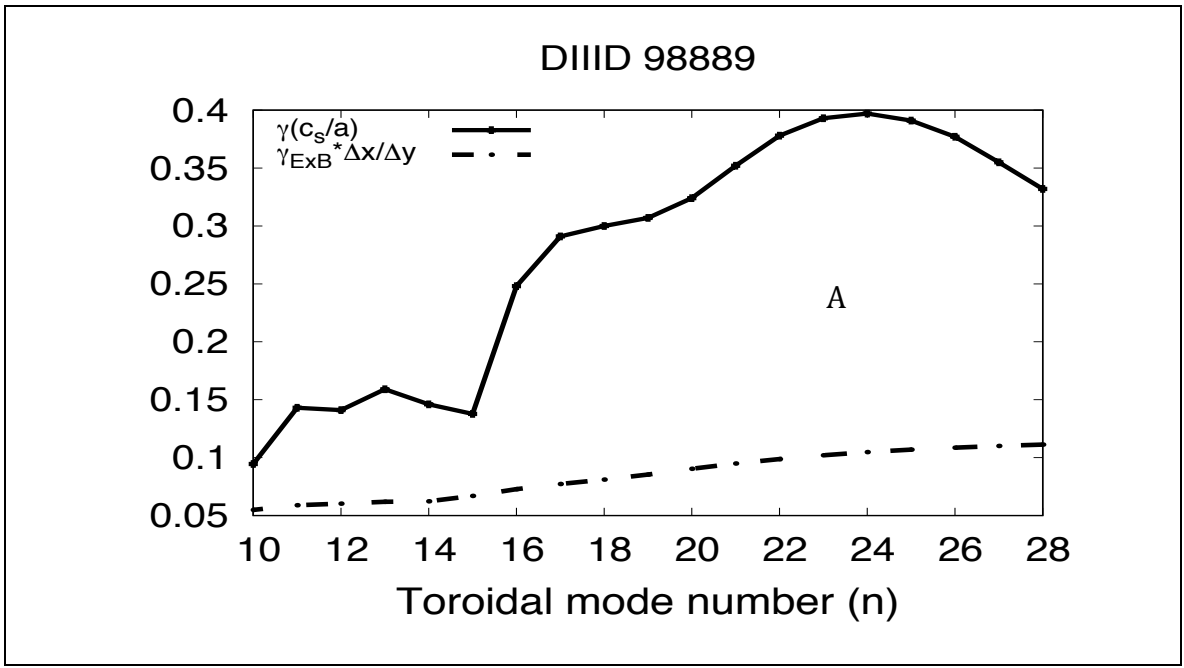

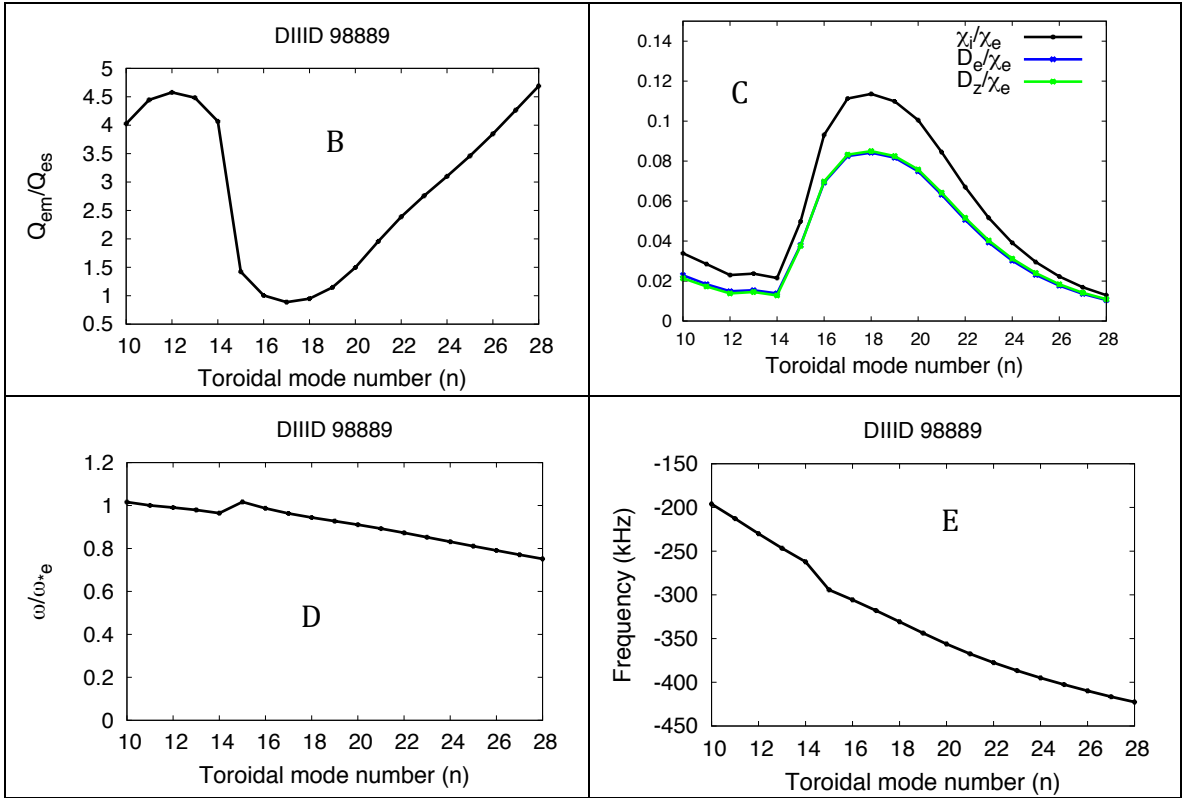

Figure 12. In (A), growth rates γ are much larger than the effective shearing rate $\gamma_{ExB}$ ($\Delta x/\Delta y$), so the modes should avoid suppression. In (B) the heat flux is mainly from magnetic fluctuations, consistent with MTM, but there can be considerable electrostatic flux. Despite some electrostatic flux, in (C), we see that the transport fingerprints are as expected for MTM In (D) the frequency is close to $\omega_e^*$, computed from the gradient of $n_e T_e$. In (E), the frequencies in the lab frame can be in the same range as the high f bands in the magnetic spectrogram.

As expected, the electron heat diffusivity is the dominant transport channel. Frequencies are close to $\omega_e^*$, and in the plasma frame, can be in the same range as the high f bands in the magnetic spectrogram (but are somewhat higher).

Local linear KBM were only found when β was increased by a large factor ~ 1.8. Thus, these modes are not likely to be unstable for this shot. However due to incompleteness in the code (mentioned earlier), we cannot, definitively, rule out the presence of an MHD-like instability. We thus use an artificially increased β as a way to examine what the characteristics of such an instability might be (and provide another test of analytical theory). The fingerprints, and overall characteristics, are in good agreement with the analytical expectations of sections V and VII.

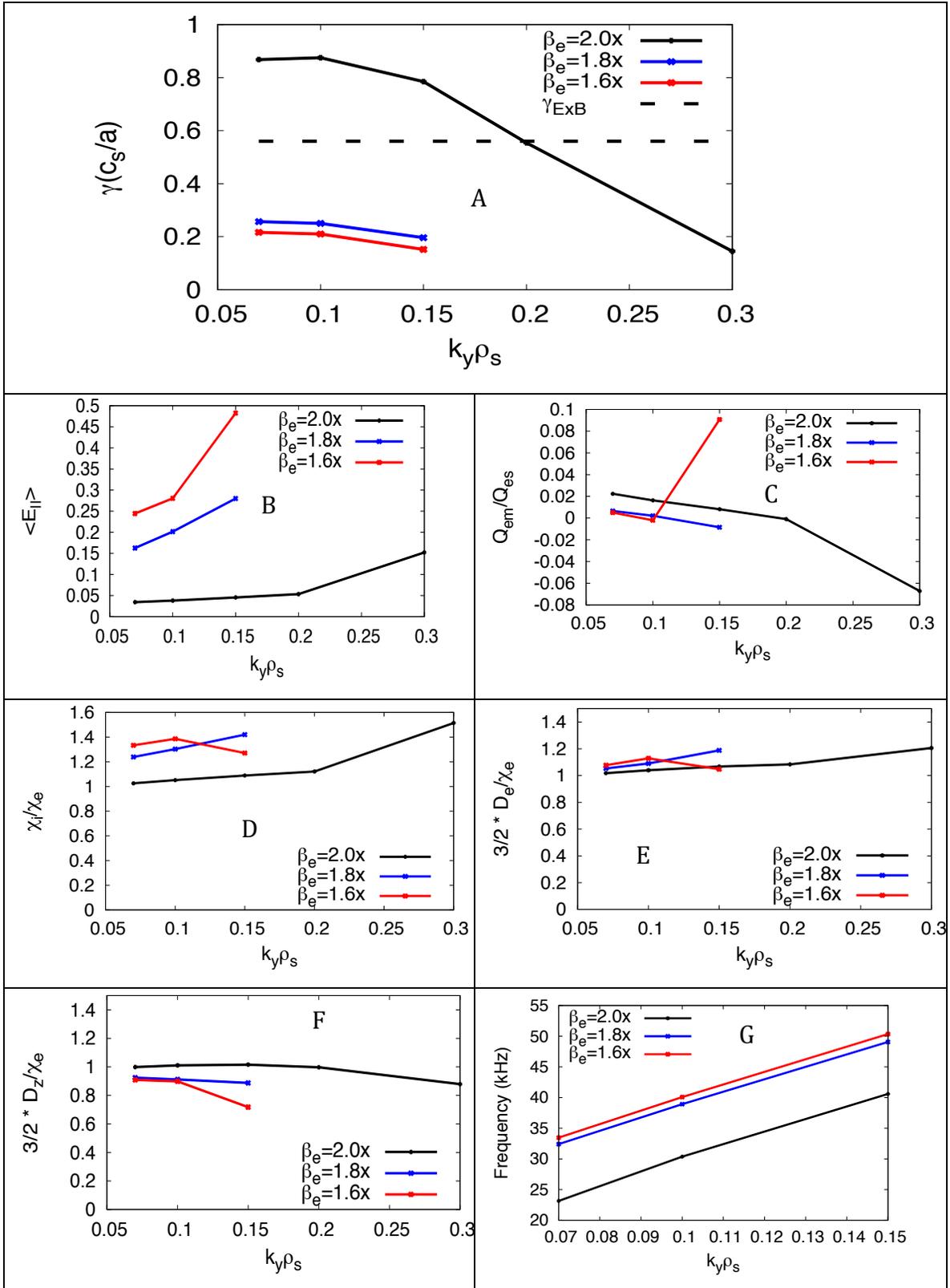

Figure 13. Characteristics for local linear KBM for DIII-D shot 98889. Missing results for $k_y\rho_s > 0.15$ and beta = 1.6x and 1.8x are not MTM and are thus not shown. In (A),

> Growth rates increase suddenly with β between 1.8x and 2x, indicative of KBM. Also, γ becomes much larger than $\gamma_{ExB}$ so the modes should avoid suppression. In (B), the average $E_\parallel$ also becomes small. In (C), the electron heat flux is mainly electrostatic, as expected for an MHD-like mode. Thus, all diffusivities are comparable, with ratios close to the analytic expectations: for $T_i$ (D), $n_e$ (E) and $n_C$ (F). In (G), frequencies in the lab frame are shown, including the measured $E_r$ Doppler shift; these are ~ 6 times less than the measured QCF.

Just like in 153674/5, the global MTM instability spectrum is limited to some discrete mode numbers n (n=16,18 and 21), separated by weakly unstable electrostatic TEM/ETG modes (see Table 5). This would tend to give quasi-coherent magnetic fluctuations, as observed. The f values for n =16 and 18 are about 1.5 times the high f bands on the magnetic spectrogram. As we will see, nonlinear simulations of these modes show a nonlinear downshift of the frequency, to within ~ 40% of the observed values.

We conclude that the high frequency QCF are much more likely to originate from an MTM rather than a KBM instability.

Table 5. Global linear MTM versus mode number n for shot 98889

| Toroidal mode number (n) | Mode type | Growth rate $\gamma(c_S/a)$ | Frequency in lab frame (kHz) |
|---|---|---|---|
| 12 | ES | 0.005 | -58 |
| 13 | ES | 0.016 | -61 |
| 14 | ES | 0.025 | -62. |
| 15 | ES | 0.035 | -64 |
| 16 | MTM | 0.18 | -291 |
| 17 | ES | 0.046 | -67 |
| 18 | MTM | 0.294 | -331 |
| 19 | ES | 0.086 | -70 |
| 20 | ES | 0.1 | -72 |
| 21 | MTM | 0.152 | -361 |
| 22 | ES | 0.124 | -76 |

Nonlinear simulation results are shown in Table 6. There is a frequency downshift, as was found in the cases for shot 152674/5. Adding the contributions from each mode, MTM can produce a total transport power ~ 2.2 MW, very close to the electron power from the transport analysis ~ 1.9 MW. The ETG power flux for this shot is ~ 0.1 MW. Thus, power balance is matches consider well within the error bars (in fact, fortuitously so, we believe).

TABLE 6 Nonlinear simulation results for profiles of DIII-D shot 98889

| Case | Toroidal number n | Heat flux at mid pedestal $\rho = 0.98$ (MW) | Linear frequency (kHz) | Nonlinear frequency (kHz) | Nonlinear frequency downshift |
| --- | --- | --- | --- | --- | --- |
| 98889 | 16 | 0.3 | -291 | -236 | 19% |
| 98889 | 18 | 1.1 | -331 | -254 | 23% |
| 98889 | 21 | 0.8 | -361 | -318 | 12% |

Linear ITG/TEM modes are found near the pedestal top. The growth rate $\gamma$ of the modes is less than the local ExB shearing rate $\gamma_E$. It has been already pointed out little transport is expected from these modes, especially in the electron thermal channel. Nonetheless, we consider their properties in section VII as an instructive example. Note that, $\chi_i/\chi_e \sim$ 3-4 for such modes.

Summary of conclusions of the experimental comparisons for shots 153674/5 and 98889

For both shots, high frequency QCF bands observed on magnetic diagnostics (and BES for 153674/5) are a good match to MTMs, but are inconsistent with KBMs.

The pedestal profile of evolution of 153674/5, in the $T_e$, $n_e$, $T_i$, and $n_C$ channels, is only consistent with the QCF being an MTM. For 98889, the transport analysis is inconsistent with KBM dominating the energy losses, for both the $n_e$ and $T_i$ channels. In both cases, a combination of MTM and ETG can produce transport power that roughly matches experiment.

Hence, we conclude that MTM and ETG are responsible for the power losses observed in these DIII-D shots.

Of particular qualitative significance, in both these cases, MTMs in the form of QCF are found to be capable of producing significant transport losses. This indicates that qualitatively similar QCF observed in many other cases (e.g. JET washboard modes [37] and high frequency QCF bands in ASDEX-U [38]), with similar gross MTM stability parameters (as indicated in table 2) are indeed strong candidates to be MTMs causing significant electron energy losses, as was inferred from the observed inter-ELM pedestal evolution for those cases. The preceding analysis is a blueprint for how to employ detailed gyrokinetic simulations in specific cases to establish this more definitely.

## Section VI: Quantitative kinetic analysis for transport fingerprints from electromagnetic modes

We now give a quantitative analysis of the fingerprints of electromagnetic modes in a pedestal. We emphasize magnetic modes, because magnetic fluctuations are frequently observed in current pedestals. The transport fingerprint of MHD-like modes, and of MTM can be obtained from fundamental theoretical considerations implicit in the drift-kinetic Maxwell system.

The Quasi-Linear Theory (QLT) has been used successfully to compute transport ratios [68-70] in the core. It is also used for the transport ratios in the widely benchmarked TGLF model for core transport [71]. There are reasons why QLT might be an even better approximation in the pedestal than in the core 1) The ETB has lower diffusivity than the core, suggesting lower fluctuation amplitudes; and 2) observed fluctuations are often quasi-coherent; both these characteristics are indicative of fluctuations closer to the linear eigenmodes than in the core. Thus, we use QLT for transport ratios in the pedestal. And importantly, QLT can describe and examine both stochastic magnetic field transport, and ExB convective transport.

We start with the drift kinetic equation (DKE) as presented by Hazeltine [72]. This reference uses a maximal ordering; the only two assumptions are that the frequency is low compared to the gyrofrequency ($\omega < \Omega$) and that the gradient scale lengths L are longer than a gyroradius ($\rho < L$) (where $\rho$ is defined using the total magnetic field B). The lack of further assumptions in this derivation is noteworthy:

> 1) No assumption about the relative sizes of frequencies other than $\Omega_I$, so this DKE can be used for both drift and MHD-like modes.
>
> 2) Typically $\rho_i/L_i$ is fairly small in a pedestal. For the DIII-D pedestals here, $\rho_i/L_i \sim 1/15 - 1/20$. Near the separatrix it is $\sim 1/3$; the modes we consider are not in this region.
>
> 3) The fluctuation can have the same scale length as the equilibrium variations, so long as $k \rho_i$ is small. No separation of scales is made between the equilibrium and the fluctuations (as is done in most derivations of the gyrokinetic equation).

We include only the lowest order terms in $\rho/L$ and $k\rho_i$ for this DKE. Furthermore, we neglect some small terms arising from magnetic field compression $\sim \delta B_{||} \sim \beta$. (Pedestals must have small $\beta$; they cannot exist far into the MHD unstable regime, so $(R/L) \beta$, which measures pressure driven stability with curvature, cannot be very large; since R/L is large, $\beta$ must be small.)

Details are similar to other calculations of kinetic instabilities, and are given in the Appendix: Details of the kinetic calculations of the fingerprint. After some algebra, linearized DKE for fluctuations can be written as:

$$-i\omega_{pl} \delta f_s + (\mathbf{v_d} + v_{||} \mathbf{b}) \cdot \nabla \delta f_s + C(\delta f_s) = \omega_s^* (q_s/T_s)(\phi + v_{||} A_{||}/c) + (q_s \delta E_{||}/T_s) f_M + (\mathbf{v_d} \cdot \nabla q_s \phi / T_s) f_M \qquad \text{Eq(6)}$$

Here, $\omega_{pl}$ is frequency $\omega$ in the local plasma frame (including $\mathbf{v_{0ExB}}$ Doppler shift), $\omega_{pl} = \omega - \omega_{ExB}(r)$ (r is a flux label), $\omega_{ExB}(r)$ and $\omega_s^*(r)$ are the general geometry versions of the ExB Doppler shift and the diamagnetic drift, and, we define $\delta E_{||} = -\mathbf{b} \cdot \nabla \delta \phi - i\omega_{pl} \delta A_{||}$, the parallel electric field as it would be in the local plasma frame.

Consideration of the convective response

It is extremely revealing to subtract out the "convective" part of $\delta f_s$; this is the response that would pertain if the perturbed distribution was determined purely by convection, $d\delta f_{conv\,s}/dt + \delta \mathbf{v_{ExB}} \cdot \nabla f_{Ms} = 0$, with $d/dt = \partial/\partial t + \delta \mathbf{v_{0ExB}} \cdot \nabla$. Simple algebra gives, for the deviation from $\partial \delta f_{conv\,s}$, $\delta f_{dev\,s} = \delta f_s - \delta f_{conv\,s}$, the kinetic equation

$$-i\omega_{pl} \delta f_{dev\,s} + (\mathbf{v_d} + v_{||} \mathbf{b}) \cdot \nabla \delta f_{dev\,s} + C(\delta f_{dev\,s}) = (1 - \omega_s^*/\omega_{pl})(q_s \delta E_{||}/T_s) v_{||} f_{Ms} + (\mathbf{v_d} \cdot \nabla q_s \phi/T_s)(\omega^*/\omega_{pl} - 1) f_{Ms} \qquad \text{Eq(7)}$$

The definition of the convective response becomes

$$\delta f_{s\,conv} = (\omega_s^*/\omega_{pl})(q_s \delta \phi/T_s) \qquad \text{Eq(8)}$$

*Equation(7) implies that deviations from purely convective response are driven by two terms: the first is proportional to ~ $\delta E_{||}$, and the second, ~ $(1/\omega_{pl}) \mathbf{v_d} \cdot \nabla$, is driven by the curvature.*

For typical modes under investigation, $\omega \sim \omega^*$. For this class of modes, the ratio of the curvature driven part of $\delta f_{dev\,s}$ to $\delta f_{s\,conv}$ is ~ $L_{ped}/R$: very small in a pedestal. *So if the $\delta E_{||}$ contribution is small, then the perturbations will be mainly convective.*

*And if the primary behavior is mainly ExB convection of all species, we expect that all channels have similar transport diffusivities, and no pinch, as we see below.* This argument is a basic corollary of the dynamics, as described in the kinetic equation, and is not strongly dependent on details of the mode structure or type.

The Quasi-Linear Diffusivity for the convective response

The convective response, inserted into the QL diffusivity, yields the velocity dependent flux for ExB convection:

$$\Gamma(v) = <Re[\delta v_{ExB}*\delta v_{ExB}/(-i\omega_{pl})]> \cdot \nabla f_0 \qquad Eq(9)$$

For a spatially dependent Maxwellian, when integrated over velocity to get the particle flux, the temperature gradient contribution vanishes, and

$$\Gamma = D \, dn_s/dr \qquad Eq(10)$$

With

$$D = <Re[\delta v_{ExB}*\delta v_{ExB}/(-i\omega_{pl})]> \qquad Eq(11)$$

There are no pinches. The heat flux has $\chi = 3/2\, D$, plus a convective energy flux :

$$Q_s = (3/2) \, D \, n_s \, dT_s/dr + (3/2)\Gamma_s T_s. \qquad Eq(12)$$

All species have the same $\chi$. These unsurprising results for convective transport will serve as a reference.

When is $\delta E_{||}$ small?

*Finally, $\omega_{pl}$ affects the size of $\delta E_{||}$; unless $\omega_{pl} \sim \omega_e^*$, as for MTM, the pedestal ordering $L/R <<1$ will imply $\delta E_{||} \sim 0$.* To see this, note quasi-neutrality, $\nabla \cdot \boldsymbol{\delta j} = 0$, can be written as $\nabla \cdot \mathbf{b}\, \delta j_{||} = - \nabla \cdot \delta j_\perp$. Note that highly mobile electrons dominate the parallel current $j_{||}$ ( in Eq(7)), whereas polarization physics dominates $j_\perp$. For electromagnetic modes in a pedestal, $\delta j_{||}$ tends to be "too large"- the tilt $\delta B$ in the direction of the strong ETB gradient gives large parallel gradient, and hence large $\nabla_{||}\delta j_{||}$, to be balanced by the polarization $\nabla \cdot \delta j_\perp$.

$$[(\omega_{pl} - \omega_e^*)/(\omega_{pl}-\omega_i^*)](\omega_e^*/\omega_{pl})^2 >> (m_i/m_e)(L_{ped}/L_s)^2 \qquad Eq(13)$$

With the simplification $\omega_{pl} \sim \omega_e^*$,

$$1 >> (m_i/m_e)(L_{ped}/L_s)^2 \qquad Eq(14)$$

For parameters in the steep gradient region of a pedestal, where scales $L_{ped}$ are small, this is usually well satisfied. But inside the pedestal top, as $L_{ped}$ begins to approach core-like gradient, this inequality will fail. Hence, in the body of the ETB, $\delta E_{||}$ will be small for modes for which $\omega_{pl} \neq \omega_e^*$; *the ideal MHD modes, KBM, and RMP belong to this class. So they will have comparable diffusivity in all channels.*

These analytic results confirm that Table 1 is an accurate description of the fingerprints for MTM and MHD-like modes, corroborating the qualitative arguments from section IV. These basic theoretical deductions were corroborated by the simulation results of section V, and also by those in the following section VII.

The analytic arguments in this section have emphasized the transport fingerprint and mode frequency. We briefly note that they also imply that another type of fingerprint can be very discriminating; the ratios of different fluctuating quantities, such as $(\delta n/n)/(\delta B/B)$, $(\delta T_e/T_e)/(\delta B/B)$ and $(\delta n/n)/(\delta T_e/T_e)$ are very different for MHD-like modes than for MTM, and this is a consequence of the same differences in underlying physics that leads to very different transport fingerprints.

For example, for MHD-like modes, eq(8), together with $0 = \delta E_{||} = i \omega_{pl} A_{||} - ik_{||} \phi$ (where parallel gradients for the MHD mode have been simplified to $-ik_{||}$) gives $(\delta n/n) = (1/k_{||} L_n) (\delta B/B)$ and $(\delta T/T) = (1/k_{||} L_T) (\delta B/B)$. What about the same ratios for MTM? The eq(7) gives $(\delta n/n) \sim (1/k_{||} L_T) (\delta B/B) (1-\omega^*/\omega_{pl}) (k_{||} v_{th}/\omega_{pl})$. This is the same size as the MHD result, but multiplied by $(1-\omega^*/\omega_{pl}) (k_{||} v_{th}/\omega_{pl})$. As we described above, the dispersion relation for MTM implies that $(1-\omega^*/\omega_{pl})$ is small, and also $(k_{||} v_{th}/\omega_{pl})$ is small when $\beta (L_s/L)^2 >> 1$, as is true in pedestals (This follows from the arguments in ref [53]; also, see the Appendix on details of the MTM and RMHD calculations). Also, the lowest order dispersion relation when $\beta (L_s/L)^2 >> 1$, implies $j_{||} \sim 0$, which further implies that $(\delta n/n)/(\delta T_e/T_e)$ is much smaller for MTM than for MHD-like modes. Of course, gyrokinetic simulations for specific profiles should be employed to corroborate this for cases where two or more fluctuating quantities are measured. Nonetheless, these analytic considerations imply that the same fundamental differences between MHD-like modes and MTM that lead to large differences in the transport fingerprints, also lead to large differences in the ratios $(\delta n/n)/(\delta B/B)$, $(\delta T_e/T_e)/(\delta B/B)$ and $(\delta n/n)/(\delta T_e/T_e)$.

**SECTION VII: Gyrokinetic analysis of DIII-D discharges 153674/5 and 98889**

Further details of the gyrokinetic simulations here clarify and strengthen the previous arguments of this paper. The local linear calculations are the staple of most pedestal simulations in the literature; in this paper, however, we also perform global linear and global nonlinear calculations. Some further computational details can be found in ref [73]

We will explore here a fundamental criterion for when the local analysis (simulations) can be trusted to reflect the essential nature of a mode. One would believe that a local theory could be representative if the simulations can safely "fit" into a box of the pedestal width. Local calculations use the lowest order ballooning transformation. It is well known that a next order calculation should, formally, be done to determine if eigenmodes are actually consistent [74,75]. Such next order corrections are virtually never applied to simulations because they entail a *very* large increase in complexity and effort. *But equilibrium quantities vary rapidly in space in a pedestal, so such considerations are important.*

The next order theories can be considered a complicated Fourier transform convolution. As a simplification, we use local simulations to find the width $\Delta k_x$ (in $k_x$) over which the mode is unstable. The product of this with the pedestal width should satisfy $\Delta k_x \, w > 2$ (as in the Heisenberg criterion) for the mode to possibly "fit" into a box pedestal width. This approximate criterion is suitable for practical calculations, and attempts to capture the qualitative essence of the next order theories. We consider it a necessary (but not sufficient) condition for an eigenmode to actually arise.

This common sense criterion rules out many of the instabilities found in local calculations. The pedestal mode that satisfies this criterion consistently is the MTM. Global linear calculations (which account for pedestal profile variations) are in agreement with this rule of thumb; local modes that fail to satisfy $\Delta k_x \, w > 2$ are not found in global calculations.

MHD-like modes

Many aspects of such modes were discussed in section V. Here we reexamine them to test whether they fit inside a box of the width of a pedestal. In figure 14, we show the growth rates vs. $k_x \rho_s$ for the KBM for the shot 153674/5. The mode is unstable only over a very small $k_x$ range. This pedestal width is $w \sim 15 \, \rho_s$. Even if $\beta$ is increased by 1.3x, $\Delta k_x \, w \sim 1$, i.e., less than 2. Hence, it apparently does not fit. *This result is borne out by global linear calculations, which fail to find a KBM in any of the pedestals considered.*

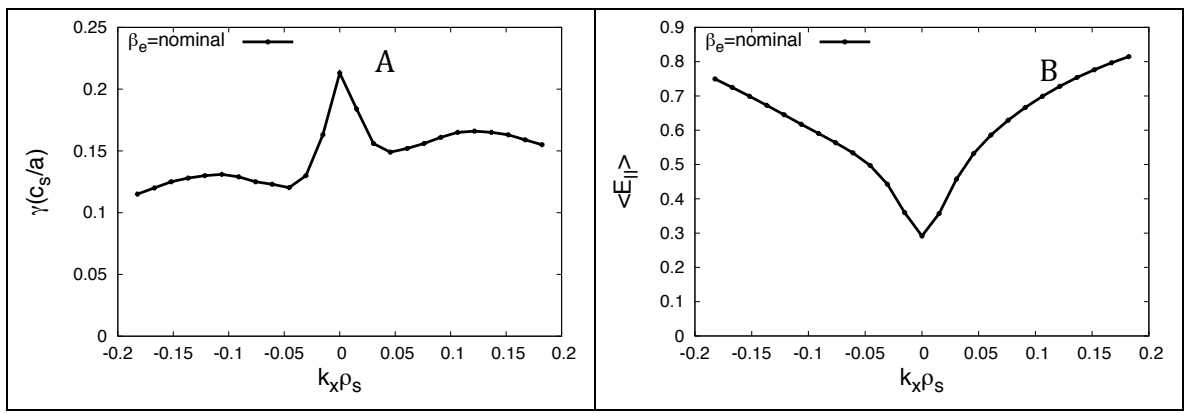

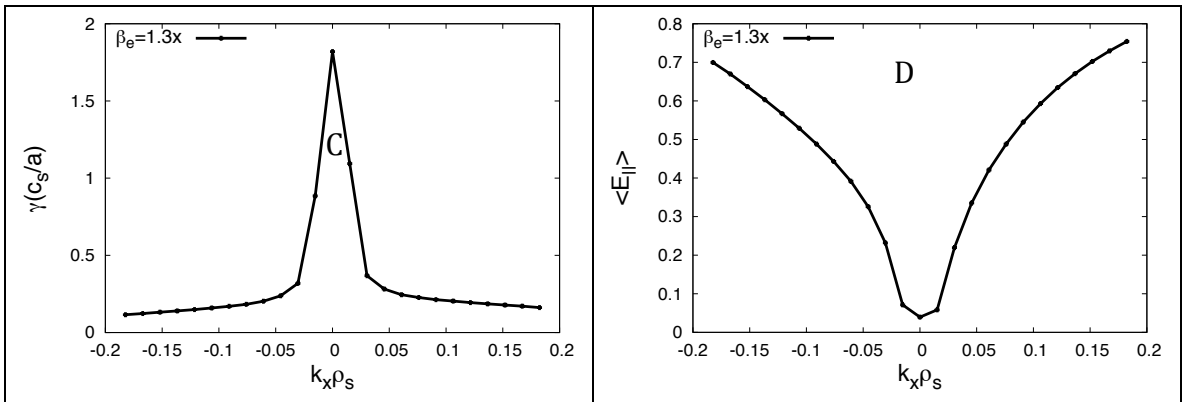

Figure 14. Growth rate vs. (normalized) radial wavenumber $k_x\rho_s$, for both the nominal $\beta$ and 1.3 times this, for the $k_y\rho_s$ of the measured fluctuation. The MHD-like mode only exists over a small range of $k_x\rho_s$, beyond which $<E_{||}>$ becomes large. So, even for $\beta$ of 1.3x nominal, the KBM cannot fit into a "box" of the width of a pedestal.

Micro-Tearing Modes

Unlike the KBM, the MTM typically have $\Delta k_x\ w \sim 4\text{-}8$, so they can easily fit into a pedestal box. Typical examples for shot 153674/5 are shown in figure 15.

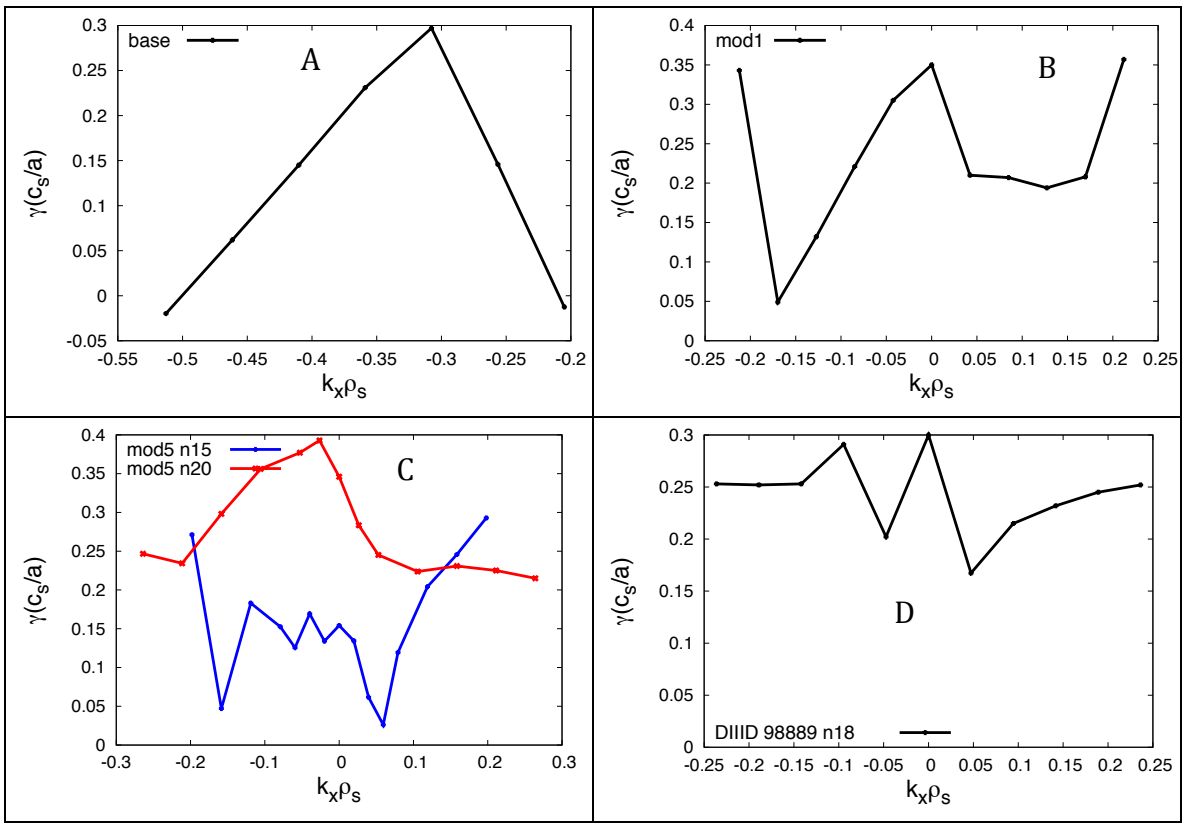

Figure 15. Growth rate vs. (normalized) radial wavenumber $k_x\rho_s$, for MTM at mid-pedestal. Cases for shot 153674/5 are shown for (A) the base case, (B) mod1, (C) mod5. In (D) we show a case for 98889 at midpedestal.

Global linear MTM simulations

Our global simulations, consistent with the arguments above, demonstrate that MTM are the only instability that exists for 153674/5 (in the range around the QCF $k_y$). And for shot 98889, MTM is by far the strongest global instability, though some weak electrostatic TEM/ETG were also found.

The MTM spectrum showed instabilities at discrete n numbers, usually separated by stable n numbers. This behavior would not have been anticipated from the local linear results, where neighboring n numbers have very similar properties. We have made substantial progress unraveling this striking feature of the global results with analytical models, but we leave such descriptions to future work.

Next, we turn to the effect of velocity shear. Local linear results in section V indicated that ExB shear should not be able to suppress these MTM. Global simulations come to qualitatively the same conclusion (Table 7).

Table 7: Global MTM for shot 153674/5 with and without ExB shear: ExB does not stabilize the modes, as indicated from local linear results.

| Case | Toroidal mode number (n) | ExB shear | Growth rate $\gamma(c_s/a)$ |
|---|---|---|---|
| mod1 | 13 | with | 0.042 |
| mod1 | 13 | without | 0.094 |
| mod1 | 15 | with | 0.067 |
| mod1 | 15 | without | 0.027 |
| mod5 | 12 | with | 0.102 |
| mod5 | 12 | without | 0.138 |
| mod5 | 15 | with | 0.252 |
| mod5 | 15 | without | 0.21 |

These global simulations include the full profile variation in the pedestal region. The outer simulation boundary was close to the separatrix ($\rho_t \sim 0.999$, where $\rho_t$ is the normalized toroidal flux.). The inner simulation boundary is chosen to fully encompass the pedestal: at $\rho_t$ = 0.95 and 0.94 for shots 153674/5 and 98889, respectively.

Global nonlinear MTM simulations

Both experimental shots find Quasi-Coherent Fluctuations. Hence, we choose nonlinear simulations that include a single n number corresponding to the QCF, as well as n=0. The simulations include nonlinear effects that are expected to be crucial saturation mechanisms for QCF:

    1) Coupling of the instability to zonal flows and GAMs
    2) Flattening of equilibrium profiles near the mode maximum, or in the vicinity of rational surfaces.

Unfortunately, nonlinear electromagnetic global GENE simulations are sometimes subject to a numerical instability; it has occurred in several unrelated nonlinear electromagnetic investigations [18,76]: after reaching a saturated state for some time, a low k numerical mode grows explosively (Fig 16). In some such cases, we were able to find an extended saturated state by slightly reducing $\beta$. The case that most closely matches experiment (mod5 for n=15) is shown on in figure 16, after reducing $\beta$ by 20%. The linear growth rate decreases by less than 20% with this reduced beta, so we do not expect that this has greatly changed the nonlinear flux. We use the result as the best available proxy for the nonlinearly saturated state of the nominal $\beta$ run.

Note that these numerical stability issues above become more serious as we try to include additional mode numbers. The dual mode number simulations presented here (n=0 plus one other n number) are the best approximation of the nonlinear physics that we have succeeded in simulating to date.

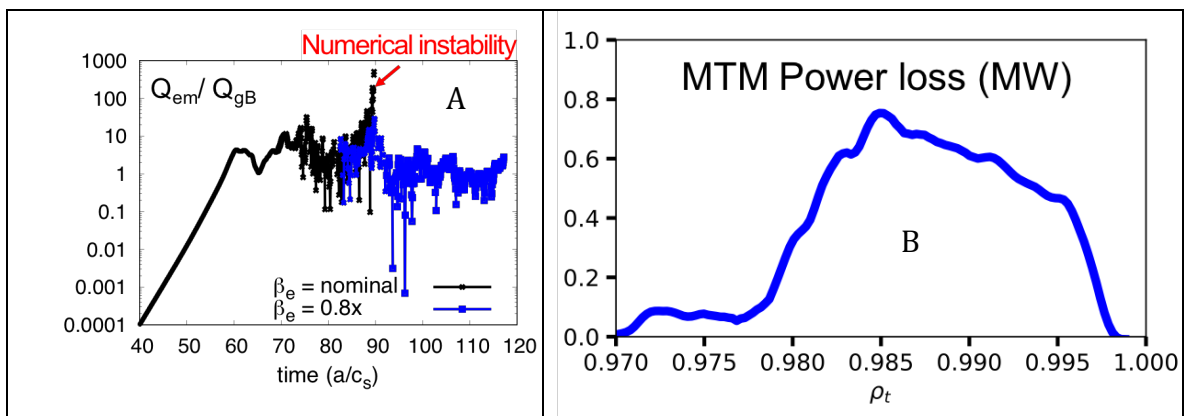

Figure 16. Nonlinear simulation results with GENE for DIIID 153764 mod5 case, n=15 are shown. The time evolution of the space averaged heat flux is shown in (A), showing the linear growth phase, nonlinear saturation phase, and nonlinear instability phase for the nominal $\beta$ simulation. Reduced $\beta$ simulation (by 0.8x)

results in an extended saturation phase. The radial flux profile in the pedestal, averaged over time, in this saturated state is in (B).

The magnitude of the maximum heat loss at reduced beta is ∼ 0.6- 0.8 MW (it is apparently higher at nominal beta). Together with heat loss from ETG ∼ 1.4 MW (Table 3), the total of heat fluxes from MTM at reduced β plus ETG is slightly above 2 MW. The total experimental power loss through the pedestal is ∼ 3 MW. Note that ELM losses and neoclassical ion heat losses are also present, in addition to MTM and ETG losses. Given the full range of experimental and simulation uncertainties, we consider this to be agreement within the (substantial) error bars.

Some of the qualitative nonlinear effects are very important. Significant profile flattening of $T_e$ occurs in the simulation, as shown in Fig.17. However, profiles of other plasma quantities are left nearly unaffected. As expected, nonlinear MTM cause mainly electron thermal transport.

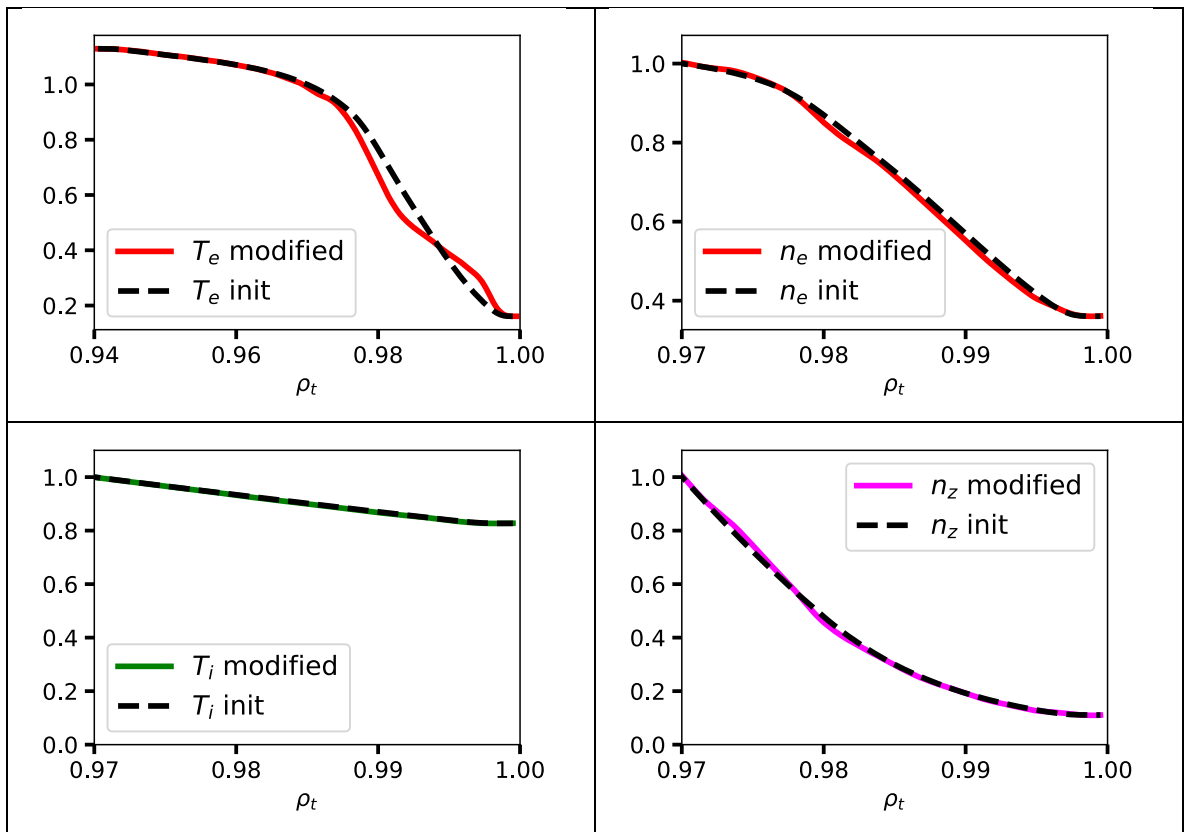

Figure 17: Profile modifications in nonlinear MTM simulations. Shown is case, mod5 n=15. Only the $T_e$ profile is significantly affected, consistent with the inter-ELM profile evolution observed experimentally on shot 153674/5. Nonlinearly modified profiles in the saturated state (solid line) are compared to the initial profiles (dashed line).

The mod 5 case shown in figure 16 -17 reached a nonlinear phase with a roughly constant heat flux. However, for some of the other modified profiles, sometimes the nonlinear simulations gave an MTM heat flux that nonlinearly crash down to very low values after reaching the nonlinear saturation state and flattening of the $T_e$ profile. This may be true even if the linear growth rate is comparable to cases with no crash. The reasons for the very different nonlinear behavior are not presently understood.

Next, we discuss global MTM simulations for shot 98889. The nonlinear runs for this shot did not collapse, and had good numerical stability, so that β did not need to be reduced. We ran nonlinear simulations for n = 16 and 18 (unstable MTM modes in the global linear simulations).  The power loss profiles calculated from the nonlinear saturation state for n = 16, n = 18, and n=21 are shown in Fig 18. Adding the power loss from the two modes together, yields a power loss of about 2.2 MW from MTM turbulence. The power loss in the electron channel found in experimental transport analysis was  ~1.8 MW, and loss from all channels was ~2.7 MW. Nonlinear ETG simulations on several locations of DIII-D 98889 (with relatively lower $\eta_e$ compared to 153674/5) has found low heat loss, about 0.1 MW. The uncertainties found for shot 153674/5 likely apply to a significant extent for shot 98889, so we conclude that MTM (primarily) can account for the anomalous power losses in this shot. The nonlinear downshift in frequency for this case was less marked than for shot 153674/5, and was in the range of ~ 10-20%.

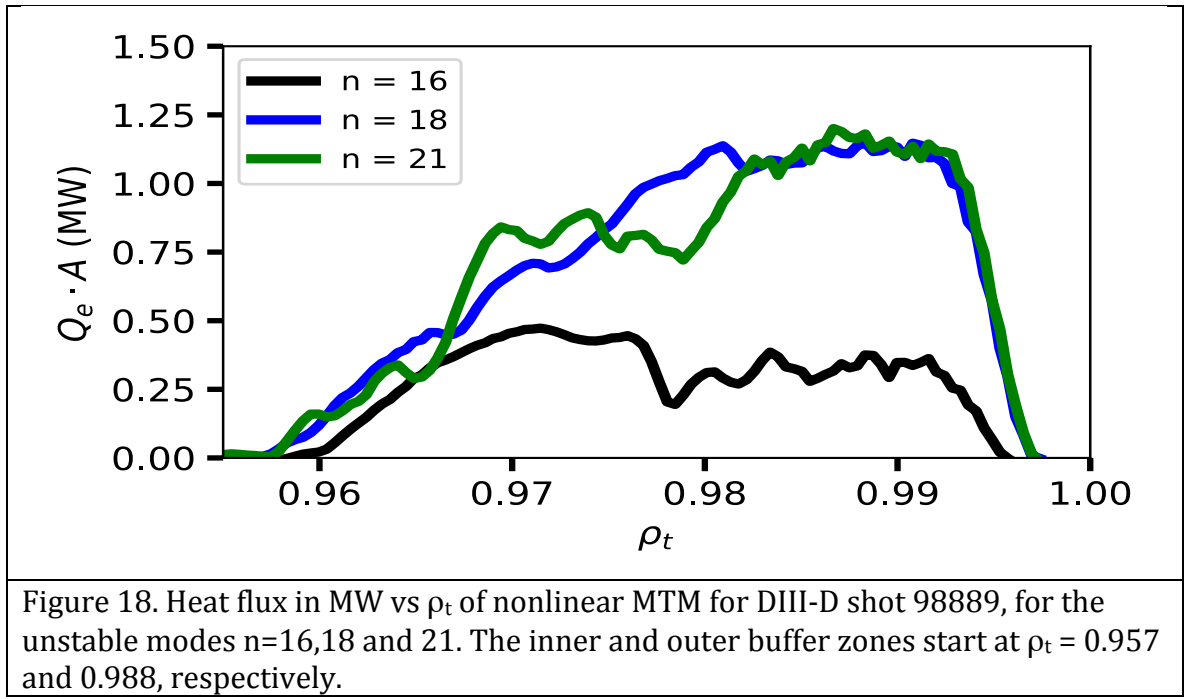

Figure 18. Heat flux in MW vs $\rho_t$ of nonlinear MTM for DIII-D shot 98889, for the unstable modes n=16,18 and 21. The inner and outer buffer zones start at $\rho_t$ = 0.957 and 0.988, respectively.

High $k_y$ ETG modes

The high $k_y$ ($k_y\rho_i > 1$) modes have already been found to produce significant transport in H-mode pedestals [15-20]. These modes have growth rates far higher than the ExB shearing rate, and the electron heat diffusivity strongly dominates the induced transport. See Figure 19.

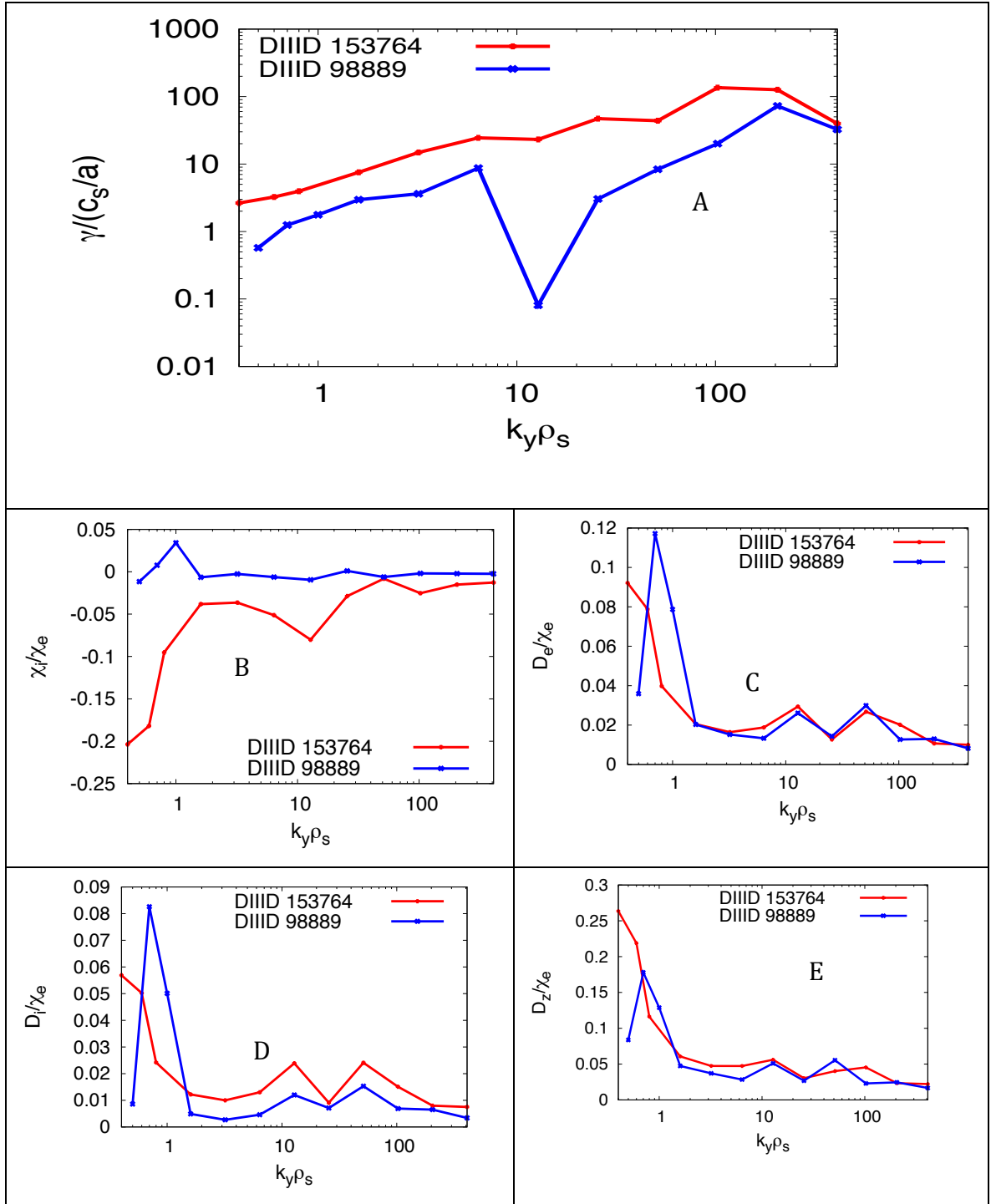

Figure 19. Linear results for ETG modes at mid pedestal. In (A) growth rates γ are much larger than the shearing rate $\gamma_{ExB}$ (which is $\sim$ 1 for both pedestals in these units). The electron thermal diffusivity dominates; the ratio of diffusivities in other channels to $\chi_e$ is small, for $T_i$ (B) $n_e$ (C), $n_i$ (D), and $n_{Carbon}$ (E).

Since the ions are nearly adiabatic, it is conventional to use local nonlinear simulations with adiabatic ions (with adiabatic ion response equal to $Z_{eff}$ $(T_e/T_i)(e\phi/T_e)$ ). Some results are shown in Table 8.

Table 8. Some details of the nonlinear ETG runs for shot 153674/5. The experimental transport power is $\sim$ 3 MW.

| Case | $\rho_t$ | $\eta_e$ | $Z_{eff}$ $(T_e/T_i)$ | $\hat{s}$ | $Q_{es}$ (WM) |
|---|---|---|---|---|---|
| base | 0.982 | 1.78 | 1.16 | 1.42 | 0.89 |
| mod1 | 0.982 | 2.40 | 1.11 | 0.76 | 4.97 |
| mod3 | 0.982 | 2.13 | 1.11 | 0.48 | 3.35 |
| mod5 | 0.982 | 1.82 | 1.13 | 0.63 | 1.39 |

The spectrum was checked to verify that it is well resolved by the spectral modes used in the simulation. As an example, see Fig.20.

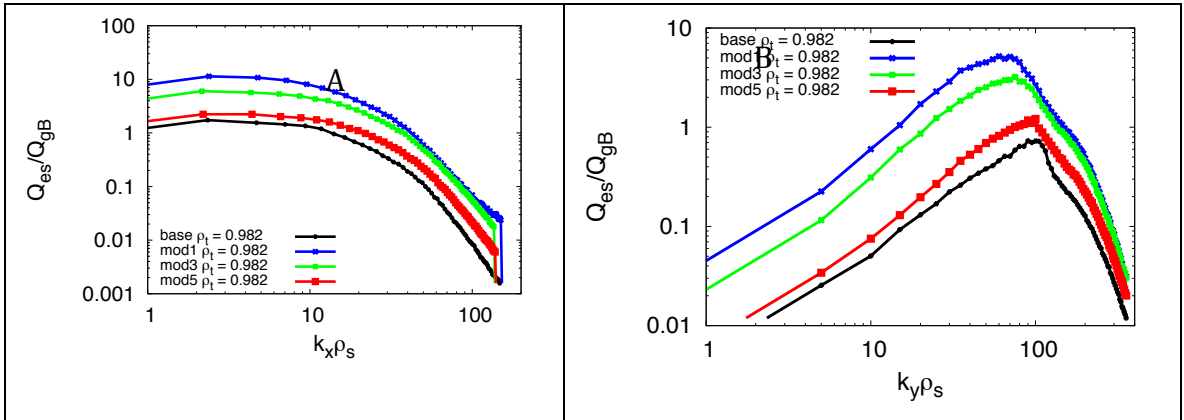

Figure 20. Typical spectrum for ETG runs are well converged (the base case is shown) in both $k_x$ (A) and $k_y$ (B)

TEM/ETG modes

In the core, the TEM modes are often the dominant instabilities at ion scales ($k_y\rho_s <$ 1). The "TEM" we find in both these pedestals are a hybrid between Trapped Electron Modes (TEM) and Electron Temperature Gradient (ETG) modes. As found previously [16], for pedestal parameters, the dominant electron driven modes for $k_y\rho_s <$ 1 can be ETG-like modes, since instabilities with large radial k, $k_r\rho_s >>$ 1 can arise. These modes can also have mode frequencies less than, or of order, the electron bounce frequency $\omega_b$. Hence, we refer to these modes as TEM/ETG modes, and they are the dominant electrostatic instabilities in both DIII-D pedestals (for most of the pedestal).

Often unstable over a broad range of $k_y\rho_s$, a representative case from 153674/5 is shown in Fig. 21 for local linear simulations near the mid pedestal.

Because of a combination of significant growth rates and small $\Delta x/\Delta y$, these modes should not be shear suppressed. However, because of the large $k_r$, the mixing length estimate of the transport $\gamma/k^2$ is extremely low- about two to three orders of magnitude lower than the thermal diffusivity estimated from power balance. Even including nonlinear deviations from the very rough mixing length estimate, these modes should not be a significant source of transport (as found in nonlinear results of ref [16]). We do not consider them further here.

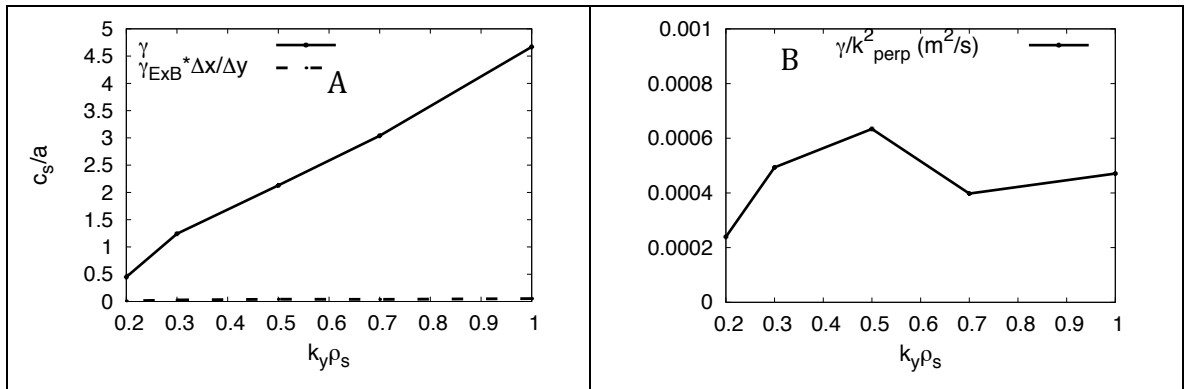

ure 21. A representative TEM/ETG mode from shot 153674/5 at mid pedestal. The modes can easily overcome shear suppression, but have a mixing length estimate of transport that is two to three orders of magnitude smaller than the estimated experimental thermal diffusivity ~ 0.2 m²/sec.

ITG/TEM

Another class of ion scale modes that play a fundamental role in determining the core transport is the Ion temperature gradient instability (ITG). For pedestal parameters, however, the ITG modes, driven by $\eta_i = d \log T_i /d \log n$, are more slab like [19,20,63]. Instabilities usually require $\eta_i \sim 1$ or more. But for shot 153674, $\eta_i$ is significantly less than one over nearly the entire pedestal. The same is close to true for shot 98889, except, at the top of the pedestal $\eta_i \sim 1$. Hence, we do not find ITG-like modes for these shots, except near the pedestal top of 98889.

Some properties of these modes are shown in fig. 22. The $\Delta x/\Delta y$ is several times less than one, which reduces their effective shearing rate $\gamma_E \Delta x / \Delta y$. Nonetheless, their growth rate is low enough so that they should still be subject to shear suppression. In addition, because of the low growth rates and relatively high $k_r$, the mixing length estimate is about two orders of magnitude lower than the estimated thermal diffusivity, as above.

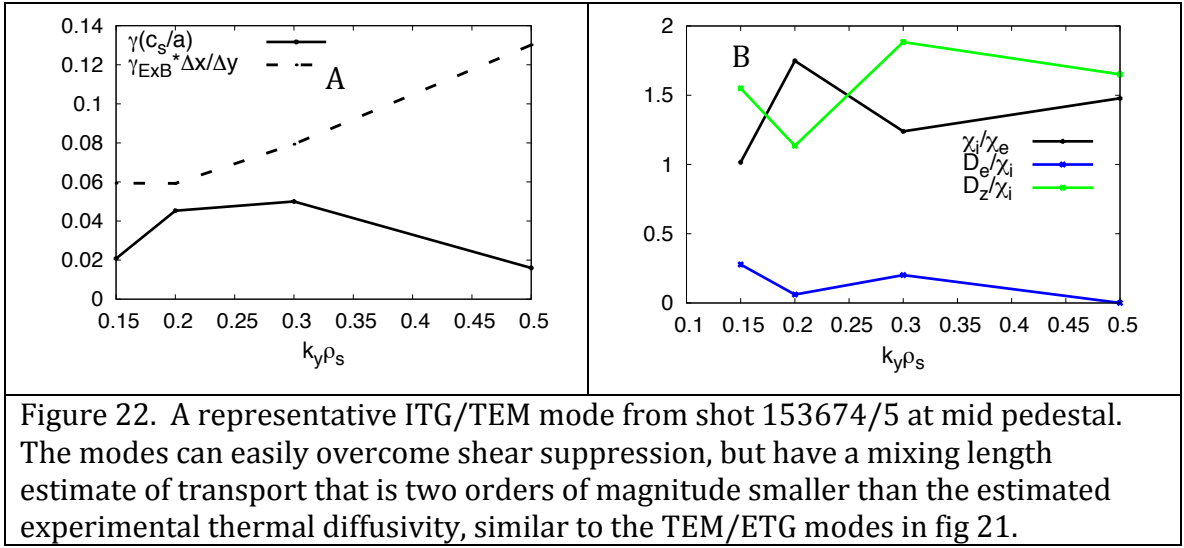

Figure 22. A representative ITG/TEM mode from shot 153674/5 at mid pedestal. The modes can easily overcome shear suppression, but have a mixing length estimate of transport that is two orders of magnitude smaller than the estimated experimental thermal diffusivity, similar to the TEM/ETG modes in fig 21.

The transport fingerprint of these modes is just as one would expect for ITG-like modes. The modes are slab-like, with $\omega \sim k_z v_{ti}$. Hence the passing electrons are adiabatic~ roughly adiabatic. Furthermore, the trapped electrons are also adiabatic, since the mode frequency is much less than the collisional de-trapping rate $\nu_{eff} \sim \nu_e/\varepsilon$, $\omega \ll \nu_{eff}$. Hence electron transport is sub-dominant. However, these modes do affect impurity particle transport (see fig 22B). As found in the analysis of the C-mod I-mode discharge [73,77], the impurity particle diffusivity can be even higher than the ion thermal diffusivity.

Alfven Eigenmode/ Micro Tearing Modes (AE/MTM)

In a pedestal, the diamagnetic frequency $\omega_e^*$ is large, and sometimes matches the frequency of an AE. We find that hybrid MTM/AE modes may appear. As variants of MTM, these modes, predominantly, produce electron thermal transport, but there is a substantial electrostatic component to transport.

The mode simultaneously satisfies $\omega \sim \omega_e^*$, and $w \sim k_z V_{Alfven}$ (where $k_z^2$ is estimated from the numerically obtained eigenfunction by $\int dz\, |d\phi/dz|^2 / \int dz\, |\phi|^2$). These modes, however, fail the test of being able to "fit" in a box with the width of a pedestal, $\Delta k_x\, w > 2$, and appear only in local linear runs for a narrow range of $k_x$.

The heat flux for these cases has a substantial electrostatic part, which can sometimes be dominant, so the mode is not a typical MTM. Also, the $E_{||}$ is small, so the mode has a significant MHD-like trait (expected for a AE). Because the ratio $\gamma/\omega$ is extremely low ($\sim 1/40$), the quasi-linear fluxes from the electrostatic fluctuations are low, as can be inferred from eq (11). The transport fingerprints of all the AE/MTM modes we find here are the same as an MTM.

In conclusion, we consider the AE/MTM to be a variant of the MTM.

**Section VIII Conclusions**

This paper, though relying heavily on nonlinear electromagnetic simulations with GENE, has added an essential new tool to investigate the dynamics that controls the pedestal transport behavior. Qualitative and quantitative analytical arguments (in sections IV and VI) extract the possible existence, the basic nature, and the defining characteristics of the instabilities responsible for the residual transport in tokamak pedestals. Subjecting the experimental data to a thorough examination, the theory, corroborated by simulations, helps us to "reconstruct", for instance, the instability(ies) that may exist in a given pedestal. The whole concept of matching the observation with the defining signatures of a mode is presented under the moniker "fingerprints". The principle fingerprints of a mode that have given us a powerful handle on understanding the physics of the pedestal transport, are: 1) the ratios of the transport diffusivities in different channels, and 2) the fluctuation frequency in the plasma frame.

A remarkably consistent picture emerges; all interpretations of disparate observations (the channels of $T_i$, $n_Z$ and $n_e$, and RMP effects), lead to the same conclusion. Some combination of MTM and/or ETG dominate energy transport in typical experimental pedestals of ELMy H-modes. MHD-like modes do not cause much energy transport, but rather, might dominate electron particle transport (and may enforce MHD marginal stability through the density evolution).

A unified explanation of all the diverse experimental observations seems to follow via the combined action of a single ansatz of a weak density source, and the theoretically inferred transport fingerprints of the modes widely expected to be responsible for transport.

This same ansatz would also allow MHD-like modes to enforce marginal stability of the inter-ELM pressure profiles via the density channel, while maintaining consistency with the wide range of experimental observations in other channels. This is apparently consistent with the inter-ELM phenomenology that pressure profiles remain close to marginal stability for a substantial fraction of the ELM cycle. It also allows the basic assumptions of the EPED model to be satisfied (wherein KBM are the specific MHD-like mode posited for the inter-ELM phase), despite the fact the energy transport is dominated by modes other than MHD-like modes.

In view of the potency and scope of the fingerprint concept, it is important to carefully establish its validity. Qualitative analytic arguments in section IV were followed by more quantitative approaches in section VI to demonstrate its generality. In section IV, general qualitative arguments clarify the nature of the modes and their properties, including how these features manifest in their fingerprints. The validity of the procedure is fortified in section VI by "deriving" the fingerprints of relevant modes from the general structure of the drift-kinetic equation within a pedestal ordering. This is corroborated by extensive gyrokinetic simulations of two DIII-D discharges in sections V and VII.

Arguments based on the transport fingerprints are indirect, in the sense that they imply that MTM and ETG must be active from their transport characteristics, rather than by direct observation. However, we also find that it is likely that MTM sometimes reveal themselves by their magnetic fluctuations; some observations on DIII-D, JET and ASDEX-U are consistent with MTM rather than MHD-like modes. The transport characteristics of JET magnetic fluctuation called "washboard modes" are only consistent with MTM modes. This is also true of similar phenomenon seen on DIII-D and ASDEX-U (the high frequency fluctuation bands). And in addition, typical parameters of ELMy H-modes are squarely in the regime where analytic theories find MTMs to be very strongly driven. Analytic theories also find that resistive MHD modes can be excluded as candidates to explain these fluctuations (for example resistive kink modes and resistive ballooning modes) because the huge diamagnetic effects in a pedestal render them to be strongly sub-dominant to the MTM.

The conclusions above are tested and illustrated by a very detailed, painstaking gyrokinetic analysis of two DIII-D shots. For DIII-D shot 153674/5, the identity of the observed magnetic fluctuation as an MTM is demonstrated by two lines of evidence: 1) the fact that the growing magnetic fluctuation produces no apparent affect on either $n_e$, $T_i$, or the impurity density $n_Z$ (implying inconsistency with an MHD-like mode), and 2) the fluctuations on this shot (and DIII-D shot 98889) must be strongly in the electron diamagnetic direction in the plasma frame (taking into account the Doppler shift from the measured $E_r$).

Both linear and nonlinear simulations of MTM and ETG were performed for both DIII-D cases. These corroborated the fingerprint analysis. And, also very importantly, for the first time, it was found that experimental levels of heat transport can be produced by Quasi-Coherent MTM of the type that are apparently observed in magnetic diagnostics. Parameter variations within the error bars show that transport levels are rather sensitive to such variations, and sensitivity studies were needed to establish the conclusion above for shot 153674/5, as well as rough agreement with the measured spectrum. In addition, ETG can be very important in matching power balance for shot 153674/5. Future gyrokinetic pedestal simulations will probably need to include such sensitivity studies to reliably reach agreement with experiments.

We believe that these results are strongly indicative that qualitatively similar QCF that are observed on other devices (e.g. washboard modes on JET and high frequency QCF on ASDEX-U) are likely important agents to transport in those devices. The examples of the application of gyrokinetic codes to the DIII-D cases here is a blueprint for similar investigations on other devices.

Nonlinear gyrokinetic simulations also revealed important qualitative nonlinear properties of the modes. In particular, there is a significant downshift of the linear mode frequency. And a QCF with a single toroidal n mode could, nonetheless, produce transport over a substantial fraction of the pedestal.

Local gyrokinetic calculations are much more widely performed within the community than global simulations, and these were also employed to reach important conclusions. At several radial locations in the pedestal, local instability frequencies, when Doppler shifted by the local $E_r$, are more consistent with the observations for MTM instabilities, not KBM. However, the calculated lab frame frequencies were still too high by a factor of about 1.5 to 2 (nonlinear simulations reduce the discrepancy considerably).

The gyrokinetic analysis of these two DIII-D pedestals verify multiple important theoretical aspects employed in the fingerprint analysis. The gyrokinetic methodology, given here, is also an example for the gyrokinetic analysis of other pedestals. In addition to considerations of the transport fingerprint and mode frequency, other important analytic concepts are used to determine the relevant instability candidates. Several novel criteria narrow the possible instabilities that need to be considered for transport:

>   a) The susceptibility to velocity shear suppression by comparison of $\gamma$ to $(\Delta x/\Delta y)\, \gamma_{ExB}$ .
>
>   b) Whether the modes found in the commonly employed local simulations are relevant in a pedestal, i.e., can they "fit" in the pedestal of width w by being unstable over a wide enough range of $\Delta k_x$ to satisfy $\Delta k_x\, w > 2$.
>
>   c) Is the very rough estimate of the mixing diffusivity several orders of magnitude too small to be significant?

Using these metrics, one finds that many of the diverse linear local instabilities are not robust enough to produce significant pedestal transport. The MTM emerge as the only robust modes with small $k_y\, \rho_i$. The ETG mode is also quite strong on shot 153674/5. However, it must be stressed that some MHD-like driving terms are not presently in the models, so such modes cannot be excluded without code improvements (that will be considered in the future).

In conclusion, identification of whether MHD-like modes, MTM and/or ETG are responsible for most "anomalous" energy losses is possible by relying on the transport fingerprints in conjunction with experimental observations of transport in multiple channels, and the application of gyrokinetic simulations guided by those concepts. In the future, ITG/TEM will be included more fully in both fingerprint analysis and gyrokinetic analysis for comparison to cases where such modes appear to be important, such as JET-ILW pedestals and possibly wide pedestal QH modes on DIII-D.

The crucial role that this work will play on deepening the understanding of current pedestals, is obvious. This is a critical stepping stone to the construction of a rather encompassing physics- based framework (tested on current machines) to analyze and help optimize pedestals for burning plasmas such as ITER.

**Acknowledgements**

This material is based upon work supported by the U.S. Department of Energy grant DE-FG02-04ER54742 and DE-AC02-09CH11466, the National Energy Research Scientific Computing Center and the Texas Advanced Computing Center.

This material is based upon work supported by the U.S. Department of Energy, Office of Science, Office of Fusion Energy Sciences, using the DIII-D National Fusion Facility, a DOE Office of Science user facility, under Awards DE-FC02-04ER54698. DIII-D data shown in this paper can be obtained in digital format by following the links at https://fusion.gat.com/global/D3D_DMP. **Disclaimer:** This report was prepared as an account of work sponsored by an agency of the United States Government.  Neither the United States Government nor any agency thereof, nor any of their employees, makes any warranty, express or implied, or assumes any legal liability or responsibility for the accuracy, completeness, or usefulness of any information, apparatus, product, or process disclosed, or represents that its use would not infringe privately owned rights.  Reference herein to any specific commercial product, process, or service by trade name, trademark, manufacturer, or otherwise does not necessarily constitute or imply its endorsement, recommendation, or favoring by the United States Government or any agency thereof.  The views and opinions of authors expressed herein do not necessarily state or reflect those of the United States Government or any agency thereof.

This work has been carried out within the framework of the EUROfusion Consortium and has received funding from the EURATOM research program 2014-2018 under grant agreement No. 633053. The views and opinions expressed herein do not necessarily reflect those of the European Commission.

**Appendix: details of the diffusivity calculation from fluid MHD**

Here we show that the ideal MHD equations for a pedestal ordering imply that the pressure perturbation evolves primarily convectively. A purely convective response in the ideal MHD equations would be a perturbation $\delta p_c$ satisfying:

$$d\ \delta p_c/dt = (\delta E \times B/B^2) \cdot \nabla p \qquad \text{eq (A1)}$$

where $\delta$ denotes perturbations and quantities without subscripts are equilibrium quantities. There is a comparable equation for the perturbed density. For the usual MHD stability calculation, done in innumerable MHD stability codes, the perturbations are around a static equilibrium.

The actual pressure evolution is:

$$d\ \delta p/dt = (\delta E \times B/B^2) \cdot \nabla p + p\ \nabla \cdot (\delta E \times B/B^2) + p\ \nabla_{||}\ \delta v_{||} \qquad \text{eq (A2)}$$

For simplicity we set the compressibility coefficient equal to 1. The first term is the perpendicular convective term as in eq A1, but there are additional perpendicular and parallel compressibility terms (second and third terms). We proceed to show these are small when L/R and $\beta$ is small.

(Note that small $\beta$ is also a corollary of large R/L; since a pedestal cannot go extremely far past the MHD stability boundary, the normalized gradient R d$\beta$/dr ~ $\beta$ (R/L) cannot be much greater than 1, hence $\beta$ must be small when R/L is large )

Consider the perpendicular compressibility term:

$$p\ \nabla \cdot (\delta E \times B/B^2) = p\ \nabla \times \delta E \cdot B/B^2 + p\ \delta E \cdot \nabla \times B\ /\ B^2 + p\ \delta E \times B \cdot \nabla B^2/B^4 \qquad \text{eq (A3)}$$

We can use $\nabla \times \delta E = \partial\ \delta B/\partial t$, $\nabla \times B = \mu_0\ j$ and $j_\perp = B \times \nabla p/B^2$. For the first term, we further use, by pressure balance, $B\partial \delta B_{||}/\partial t \sim \mu_0\ \partial \delta p/\partial t$, and also, $\nabla_{||} \sim (L/R)\ \nabla_\perp$. Compared to the convective term in eq A2, it is relatively small by two small factors, (L/R) $\beta$ (note that this is not the gradient of $\beta$, but $\beta$ itself, which enters here, and this is of the order of a percent). The second term is small by ~ $\beta$, and the third term is small by ~ L/R. Hence, the perpendicular compressibility is small compared to the convective term.

We now examine the parallel component of the momentum equation to determine the last term in eq A2.

$$-i\omega\ \rho\ \delta v_{||} = -\nabla_{||}\ \delta p \qquad \text{eq (A4)}$$

When coupled to eq (A2), parallel sound wave physics results. Hence, the last term in eq a2 is

$$p \nabla_{||} \delta v_{||} \sim \sim (p/\omega \rho) \nabla_{||}^2 \delta p \sim c_s^2 k_{||}^2/\omega \ \delta p \qquad \text{eq (A5)}$$

Even if this term is not small, it will lead to similar effects in the density equation and the temperature equation. And the kinetic analysis of section V fully includes these effects. However, within a fluid treatment, this term is small relative to the left hand side by

$$(c_s k_{||}/\omega)^2 \ll 1 \qquad \text{eq (A6)}$$

The smallness of this requires a quite low frequency or growth rate, in comparison to other relevant frequencies:

> 1) The ExB shearing rate. This is typically an order of magnitude larger than $c_s k_{||}$, estimating $k_{||} \sim 1/qR$. Estimating $E_r$ from pressure balance, $e\,n\,E_r \sim dp/dr$, then the shear of the ExB velocity is $\sim (\rho_i/L)(v_i/L)$, and $c_s k_{||}/\gamma_E \sim (L/qR)(L/\rho_i)$. Although $L/\rho_i \sim 210\text{-}20$, $L/R \sim 10^{-2}$ and $L/qR \sim 3 \times 10^{-3}$ is far smaller. Hence an instability that is strong enough to avoid shear suppression satisfies eq A6.

> 2) The diamagnetic frequency. In the presence of strong diamagnetic effects, the mode $\omega \sim \omega^*$ (in the simplest fluid treatments, $\omega \sim \omega_i^*/2$). Using $\omega^* \sim k_\theta \rho_i v_i/L$, $k_\theta \sim qn/a$, $c_s k_{||}/\omega^* \sim (L/R)(a/\rho_i)(1/nq^2)$, $(a/\rho_i) \sim 300$, $1/q^2 \sim 1/10$, so this is $\sim (L/R)(30/n)$. This is small, especially for n significantly above 1, such as is usually observed or simulated ($n \sim 5\text{-}20$).

**Appendix: details of the MTM and Fluid Resistive MHD calculations**

Consider fluid resistive modes. We estimate their growth rate for the maximum possible MHD drive, before the mode becomes an ideal MHD instability. For this case, it turns out that all the fluid RMHD modes have very similar growth rates.

For resistive kink modes driven by current gradients [50,55], the dispersion relation is

$$\omega (\omega-\omega_i^*)(\omega-\omega_e^*) = i\,C\,(v_A/L_s)^2\,(k_y^2\,D_M)$$

where $D_M$ is the magnetic diffusion coefficient ($\eta_r c^2/4\pi$ in cgs units, where $\eta_r$ is the plasma resistivity). The constant $C = 5/2$ for this case.

For resistive interchange modes the dispersion relation is the same [56], but for the maximum drive before an MHD instability occurs, $C = \frac{1}{2}$. For resistive ballooning modes of the type considered in ref [54], the coefficient C is $\frac{1}{2}\,\alpha^2/\hat{s}^2$, with $\alpha=$

$q^2 R\beta/L$. For the s-$\alpha$ model used to derive this, the ideal MHD stability boundary is at $\alpha \sim 0.7$ for $\hat{s} = 1$, so $C \sim \frac{1}{4}$. For the resistive ballooning mode considered in ref [78], the value of C is the same as for the kink mode case above. The maximum value of C for all these cases is $C = 5/2$, so we use this as an upper bound on the growth rate of the fluid RMHD modes, for MHD driving just before an ideal MHD instability arises. To evaluate figure 3 and Table 2, we take $|\omega_i^*| \sim |\omega_e^*|$ for simplicity.

To estimate pedestal gradients, we use the correlation for the pedestal width $w = 0.76 \beta_{pol}^{1/2}$[4,5]. Since this is the average width for n and T, we estimate the pressure gradient for both n and T in $dp/dx = (T\, dn/dx + n\, dT/dx)$. We use this to evaluate $\omega^*$. The $k_y$ on the outboard midplane estimated from the n number starting from the exact relation $k_y = (n/R)(B_{toroidal}/B_{poloidal})$, and $B_{poloidal}$ in the large aspect ratio limit assuming most current is concentrated on axis, so $2\pi a_{min} B_{poloidal} = \mu_0 I_p$. The collision frequency at mid-pedestal is taken as about twice the value at the pedestal top, and is adjusted for $Z_{eff}$. The Shafranov shift compression on the outer midplane is taken as $\sim 25\%$. In the cases where actual profiles and equilibrium are available, all these estimates gave reasonable agreement for the final answer (within a factor of plus or minus 1.5).

Despite the roughness of the estimates, the qualitative conclusions we draw from the order of magnitude estimates in table 2 are unlikely to be changed by the approximations. Probably the main uncertainty is the value of the magnetic shear $\hat{s} = d\log q / d\log r$ (where r is a radial coordinate). We have taken $\hat{s} = 1$, whereas numerical MHD equilibria of various pedestals find that it could be as large as $\sim 3$ or as small as $\sim 0$. If $\hat{s}$ is in the interval [0,3] the growth rates in the fourth column, $\gamma_{FRMHD}/\omega^*$, could be changed by a factor in the interval [0,10]. Hence they would still be very small compared to the MTM growth rates. The column $\gamma_{0F}/\omega^*$ might be changed by about a factor between [0,2], and hence diamagnetic effects are still very strongly stabilizing to fluid RMHD modes. And $\hat{\beta}$ could be changed by a factor in the interval [0.1,infinity], so it would still be much larger than 1. Other uncertainties likely have less effect than this. Hence, the qualitative conclusions we can draw from this table still hold.

The lowest order dispersion relation for MTM can be found by examining the size of terms in Ampere's Law, which governs the magnetic fluctuations. In the following we give a brief heuristic description of the analysis first performed in ref [53]. Neglecting $\phi$ (roughly valid for pedestal MTMs), and for the sheared slab model, this becomes

$$d^2 A_{||}/dx^2 = (4\pi/c) j_{||} = (4\pi/c)\, \sigma(x)\, A_{||} \sim (1/\lambda_{skin}^2)\, A_{||}$$

where $\sigma(x)$ is the kinetically derived conductivity, and the magnitude of the LHS is $\sim 1/\lambda_{skin}^2$ around the rational surface. Also, $\sigma(x)$ is localized around a rational surface with a width $\Delta x$ (defined by when $k_{||} v_e < (\omega \nu_e)^{1/2}$ [51,52]). Note $\Delta x^2/\lambda_{skin}^2 \sim \hat{\beta} = \beta_e(L_s/L)^2$. When $\hat{\beta} \gg 1$, the LHS is too large to be balance by the RHS, so the lowest order dispersion relation is $\sigma = 0$ at the rational surface. We use the expressions for $\sigma$ in the references and solve the resulting equation for $\omega$ numerically to obtain the results in figure 3 (assuming a value of $\eta_e = 2$, typical in pedestals).

**Appendix: details of the kinetic calculations of the fingerprint**

$-i\omega\, \delta f + (\mathbf{v_d} + \mathbf{v_{0ExB}} + v_{\|}\, \mathbf{b}) \cdot \nabla\, \delta f + C(\delta f) = (\delta \mathbf{v_{ExB}} + v_{\|}\, \delta \mathbf{b}) \cdot \nabla f_0 + q(v_{\|}\delta E_{\|} - \mathbf{v_d} \cdot \nabla \delta\phi)\, \partial f_0 / \partial \varepsilon$

Eq(A6)

Notations are standard; $\delta$ refers to fluctuations, subscript 0 denotes equilibrium quantities; $v_{\|}$ and $\mathbf{v_d}$ are the parallel and drift velocity, and $\mathbf{b} = \mathbf{B}/|B|$.

All of the modes posited as candidates for pedestal transport are flute-like (e.g., MHD-like, KBM, MTM, ITG, ETG, TEM, etc). Hence, we use $\nabla_{\|} \ll \nabla_{\perp}$. This allows us to simplify the evaluation of perpendicular gradients; we can use $B_t\, \partial/\partial \zeta \approx B_p\, \partial/\partial \theta$.

We start with conventional assumptions that $f_0$ is a local Maxwellian, and $\phi_0(r)$, i.e., the potential is a constant on a flux surface. We will discuss more complex cases later. We now discuss the form of the various terms in eq (A1). First consider the term from convection in the equilibrium $E_r$, which can be written as local Doppler shift:

$\mathbf{v_{0ExB}} \cdot \nabla\, \delta f = i\omega_{ExB}(r)\, \delta f$ $\qquad$ Eq(A7)

where $\omega_{ExB}(r)\, n = n\, (\partial \phi_0 / \partial r)/(\partial \psi / \partial r)$, with $\psi$ the poloidal flux function, and $n$ the toroidal mode number.

For the terms from the perturbed ExB drifts,

$\mathbf{B} \times \nabla\, \delta\phi / B^2 \cdot \nabla f_0 = i\, \omega_s^*(r,v)\, (q_s\phi/T_s)\, f_{Ms}$ $\qquad$ Eq(8)

where the diamagnetic frequency is

$\omega_s^*(r,v) = n\, [(cT_s/q_s)/\partial \psi / \partial r]\, (\, [\,(\partial n_s/\partial r)/n_s + (\partial T_s/\partial r)/T_s\, (\varepsilon/T_s - 3/2)]$ $\quad$ Eq(9)

The $\omega_{ExB}(r)$ and $\omega_s^*(r)$ are the general geometry versions of the perhaps more familiar expressions for large aspect ratio circular flux surfaces, $\omega_{ExB}(r) = k_y\, \mathbf{v_{\theta ExB}}$ and $\omega_s^*(r) = k_y\, (cT_s/q_sB)\, )[\, (\partial n_s/\partial r)/n_s + (\partial T_s/\partial r)/T_s\, (\varepsilon/T_s - 3/2)]$, with $k_y = n\, q_{safety}/r$, for minor radius $r$ and safety factor $q_{safety}$.

Terms from the magnetic perturbations, from $\delta A_{\|}$ proceed similarly.

The basic theoretical structure here is also robust to deviations of the equilibrium distribution function from a Maxwellian as a function of r. Pitch angle dependencies of $f_0$, $f_0(r,\varepsilon,\mu)$, (which lead to variations of physical quantities in the poloidal angle) are easily included in the formalism here by a simple redefinition of $f_M$ and of $\omega_s^*(r,\varepsilon,\mu)$ to pitch angle dependent versions, without affecting the structure of the equations (and hence the essential results). This type of distribution function would pertain, to lowest order, in the low collision frequency limit, even if finite poloidal gyroradius effects were included.

Even more general ion distribution functions can be handled, such as those that arise with both finite poloidal gyroradius and significant collisions. A general axisymmetric ion equilibrium distribution function, would be an arbitrary function of all variables $f_0(r,\theta,\varepsilon,\mu)$. It is nonetheless true that the ion response is mainly convective, because $\omega^*$ is $\gg v_i \nabla_{||}$. Using the conventional estimate of the latter as $1/q_{safety}R$, the relative size of the parallel term is $\sim L_{ped}/(q_{safety}R\, k_y\rho_i)$ which is usually quite small. Then, the ion response is always the convective one. The electrons have much smaller gyroradius, and higher collision frequency, so the distribution function can be taken to be a local Maxwellian for them. In this case the essential results described above continue to hold. *The electrons will also be primarily convective if $\delta E_{||}$ is small*; in such a system even electron diffusivities are all equal.

The Quasi-Linear Diffusivity for the convective response

Note that the QL diffusivity, D in eq(15) is proportional to the imaginary part of $1/\omega_{pl} = \gamma/Re(/\omega_{pl})^2$ (where $\gamma = Im(\omega)$). Thus its magnitude is smaller for modes with small $\gamma/\omega_{pl}$. When there is a spatial dependence of $\omega_{pl}$ due to the $E_r$ well, the real part of the denominator vanishes, but the spatial average gives a finite contribution in the usual way for resonance integrals.

When is $\delta E_{||}$ small

We can use the so called shear Alfven Law, also called the vorticity equation in MHD. Linearizing and neglecting the curvature term (which can be estimated to be small here):

$$\nabla \cdot [\, \mathbf{B} \times (\rho\, d\mathbf{v}/dt - \nabla\cdot\pi)\, /B^2\,] = \nabla \cdot \mathbf{b}\, \delta j_{||} \qquad\qquad Eq(17)$$

This equation is derived from essentially exact fluid moments, and quasi-neutrality $\nabla \cdot \mathbf{b}\, \delta j_{||} = -\nabla \cdot \delta j_\perp$, using $\delta j_\perp$ computed from the momentum equation.

For our purposes, we need only to estimate the magnitude of both sides of Eq(17). The flows are mainly ExB and diamagnetic, so the LHS $\sim (\omega - \omega_i^*)\, (n\, m_i/B^2)\, \nabla^2 \delta\phi$.

From the point of view of the transport fingerprint, the most important deviations from an MHD-like fingerprint arise due to magnetic electron transport. This requires the electrons to not be adiabatic. This allows us to restrict attention to the region near a rational surface, and arrive at analytic estimates. Near such a surface, we can estimate $k_{||} \sim k_y \Delta r/L_s$, where magnetic $L_s$ is the shear length. The relevant non-adiabatic region $\omega_{pl} \sim v_e k_{||}$, which gives $\Delta r \sim (\omega_{pl}/\omega_e^*) \rho_s (m_e/m_i)^{1/2}(L_s/L)$. Pedestals are typically in a kinetic regime where the inertia is, at least, comparable to collisions, so $j_{||} \sim (ne^2/m_e)(\omega_{pl} - \omega_s^*)(\omega_{pl})^{-2} \delta E_{||}$. (collisions can be included, but will not affect the ultimate criterion). We can bound the perpendicular gradients as $\sim 1/\Delta r$ (the largest possible size, and hence, largest possible $\delta j_\perp$). Then the dominant terms in eq(17) (with polarization on the LHS and $\delta j_{||}$ on the RHS) become:

$$(\omega_{pl} - \omega_i^*)(n m_i/B^2) \delta\phi/\Delta r^2 \sim (ne^2/m_e)(\omega_{pl} - \omega_s^*)(\omega_{pl})^{-2} k_{||}(k_{||} \delta\phi - i \omega_{pl} \delta A_{||})$$
Eq(18)

When the LHS is relatively small, polarization physics is incapable of balancing the tendency of $\delta j_{||}$ to cause a charge separation. In such a case, Eq(18) implies that $\delta\phi$ must give an electrostatic $\delta E_{||}$ to cancel the inductive part of $\delta E_{||}$, in order to reduce $\delta j_{||}$. Specifically, this occurs when the coefficient of $\delta\phi$ on the LHS is smaller than the coefficient of $\delta\phi$ on the RHS. Using the expression for $\Delta r$, we obtain that $\delta E_{||}$ is small when $[(\omega_{pl} - \omega_e^*)/(\omega_{pl}+\omega_i^*)](\omega_e^*/\omega_{pl})^2 >> (m_i/m_e)(L_{ped}/L_s)^2$, as given in the body of the paper.